\documentclass[a4paper]{article}
\usepackage{graphicx}
\input xepsf

\begin {document}
\title {\bf 3D Quantum Trajectories. Quantum orbits of the Hydrogen's electron}
\author{ T.~Djama\footnote{Electronic address:
{\tt djamatoufik@yahoo.fr}}}
\date{\today}
\maketitle

\begin{abstract}
\noindent In this paper, we introduced
the 3D-Quantum Stationary Hamilton Jacobi Equation for a central
potential, and established the 3D quantum law of motion of an
electron in the presence of such a potential. We established a
system of three differential equations from which, and as a
numerical application, we plotted the 3D quantum trajectories of the
Hydrogen's electron for the ground state and two excited states. we
did a comparison between a supposed classical trajectory of
the hydrogen electron and the corresponding QTs. We
found out that for a particular set of the hidden variables, the
QT is close in its shape to the classical one.
\end{abstract}
\vskip\baselineskip

\newpage
\vskip0.5\baselineskip
\noindent
{\bf 1.\ \ Introduction }
\vskip0.5\baselineskip
More than five years ago, we presented a deterministic approach of quantum
mechanics in 3D [1] based on the 3D Quantum Stationary Hamilton Jacobi Equation
(3D-QSHJE)
\begin{eqnarray}\label{continuity}
{1 \over 2m_0}\left({\vec{\nabla}S_0(\vec r)}\right)^2-{\hbar^2
\over 2m_0}{\Delta A(\vec r) \over A(\vec r)}+(E-V(\vec r))=0\; \;
, \hskip15mm  \label{qshje} \\
\vec{\nabla} \cdot \left(A^2(\vec r)\vec{\nabla}S_0\right)=0 \; \; . \hskip35mm
\end{eqnarray}
In that work, we established the solution of the 3D-QSHJE to be of the form
\begin{equation}\label{action}
S_0(\vec{r})=\hbar \arctan\left({\Psi_1(\vec{r}) \over
\Psi_2(\vec{r})}\right) \; ,
\end{equation}
where $\Psi_1$ and $\Psi_2$ are two real independent solutions
of the 3D Schr{\"o}dinger equation (SE).
The expression of the function $A(\vec r)$ is given as
\begin{equation}
A(\vec{r})=\sqrt{\Psi_1^2+\Psi_2^2}\; .
\end{equation}
In addition, we established the 3D quantum law of motion \cite{Dja1}
\begin{equation}\label{3d-qlmotion}
\vec{v}\; \cdot \; \vec{\nabla S_0}=2\left[E-V(\vec{r})\right]\; .
\end{equation}
This last equation is the one that should be be used to plot the
3D quantum trajectories (QTs).

The establishment of the 3D quantum law of motion
(Eq.~(\ref{3d-qlmotion})) for a general form of the potential is
an important step to investigate the 3D motions. However, we
should plot the 3D QTs when it is possible in
order to understand the behavior of the quantum particle's motion in 3D
spaces. With this aim, we should proceed exactly as we have done for the
1D quantum and relativistic QTs \cite
{Dja2,Dja3}. Indeed, we must first solve the 3D SE to find the functions $\Psi_1$ and
$\Psi_2$, which we take into the expression of the reduced action
(Eq. \ref{action}). Then, we take expression (\ref{action}) into
Eq.~(\ref{3d-qlmotion}). After solving Eq.~(\ref{3d-qlmotion}), we
can plot the QTs. Note that for a general
potential, plotting the 3D QTs is not obvious
due to the complexity of the problem.
This is why, in this article, we study the case of the central
potential, and plot the trajectories of an electron moving under
the atomic potential.

The paper is organized as follows: In Sec.
2, we introduce the 3D-QSHJE and its solutions for the central
potential. Then, in Sec. 3, we rewrite the quantum law of motion
for this potential and establish a system of three differential
equations that describe the dynamical behavior of the quantum
particles. After that, in Sec. 4, we plot the classical trajectories of a particle moving
under the action of the atomic potential.
In Sec. 5, we plot the 3D QTs of the Hydrogen's electron for several bound states. Finally, In Sec. 6,
we present a conclusion in which we discuss our results.

\vskip\baselineskip \noindent {\bf 2.\ \ The 3D-QSHJE for the
central potential} \vskip0.5\baselineskip
In order to construct the 3D-QSHJE in the case of
a central potential, we write the 3D SE with respect to the spherical coordinates
$(r,\vartheta,\varphi)$ \cite {Coh1}
\begin{equation}\label{spher-schro}
{\partial ^2 \Psi \over \partial r^2}+ {2 \over r}{\partial  \Psi
\over \partial r}+ {1 \over r^2}{\partial ^2 \Psi \over \partial
\vartheta^2} +{\cot \vartheta \over r^2}\,{\partial  \Psi \over
\partial \vartheta}+ {1 \over r^2 \sin^{2} \vartheta }\,{\partial
^2 \Psi \over \partial \varphi^2}+ {2m_0 \over \hbar^2}[E-V(r)]
\Psi=0 \, ,
\end{equation}
$\Psi(r,\vartheta,\varphi)$ is the wave function of the considered
quantum system. To solve Eq. (\ref{spher-schro}), the variables
must be separated into $\Psi(r,\vartheta,\varphi)$ as \cite {Coh1}
$$
\Psi (r, \vartheta, \varphi)=R(r)\, T(\vartheta) F(\varphi)\ ,
$$
where $R(r)$ is the radial wave function, $T(\vartheta )$ and $F(\varphi
)$ are the angular wave functions. It follows that Eq.
(\ref{spher-schro}) will be separated into three equations, each
one corresponds to one variable,
\begin{equation}\label{1d-schro-radial}
{\frac{r^{2}}{R}}{\frac{d^{2}R}{dr^{2}}}+{\frac{2r}{R}}{\frac{dR}{dr}}+{%
\frac{2m_0\, r^{2}}{\hbar ^{2}}}(E-V(r))=\, l(l+1) \;,
\end{equation}
\begin{equation}\label{1d-schro-theta}
{\frac{d^{2}T}{d\vartheta ^{2}}}+\cot \vartheta {\frac{dT}{d\vartheta }}%
+\left( l(l+1) -{\frac{m^{2}_{l}}{\sin ^{2}\vartheta }}\right) T=0\;,
\end{equation}
\begin{equation}\label{1d-schro-phi}
{\frac{d^{2}F}{d\varphi ^{2}}}+m^2_{l}\ F=0\; .
\end{equation}
$l$ being a positive integer or nil, and $m_{l}$ is an integer
satisfying to
$$
-l \leq m_{l}\leq {l}\;.
$$
$l(l +1)$ is the eingen value of $L^{2}$, $L$ representing the
angular momentum operator. $m_{l}$ represents the eingen value of
$L_{z}$, the operator projection of $L$ along the $z$ axis. Now,
let us write Eqs.~(\ref{1d-schro-radial}), (\ref{1d-schro-theta})
and (\ref{1d-schro-phi}) in a form analogous to the 1D SE. First, we take up
Eq.~(\ref{1d-schro-radial}) and write
\begin{equation}
R(r)={\frac{\mathcal{X}(r)}{r}}\;,
\end{equation}
where $\mathcal{X}(r)$ is a function of the radius. This will
allow us to write
\begin{equation}\label{pseudo-schro-rad}
{\frac{-\hbar ^{2}}{2m_0}}{\frac{d^{2}\mathcal{X}}{dr^{2}}}+\left[ V(r)+{\frac{%
\lambda \hbar ^{2}}{2m_0r^{2}}}\right] \mathcal{X}=E\mathcal{X}\ .
\end{equation}
This last equation has the same form as the 1D SE with a fictitious potential
\begin{equation}
V^{^{\prime }}(r)=V(r)+{\frac{\lambda \hbar ^{2}}{2m_0r^{2}}}\ .
\end{equation}
Likewise, Eq.~(\ref{1d-schro-theta}) can be reduced to the form of
the 1D SE by introducing the function
\begin{equation}
\mathcal{T}(\vartheta )=\sin ^{\frac{1}{2}}\vartheta T(\vartheta )
\end{equation}
into Eq.~(\ref{1d-schro-theta}), we will find
\begin{equation}\label{pseudo-schro-theta}
{\frac{d^{2}\mathcal{T}}{d\vartheta ^{2}}}+(\lambda +{\frac{1}{4}})\mathcal{T%
}+{\frac{(1/4-m_{l}^{2})}{\sin ^{2}\vartheta }}\mathcal{T}=0\
.
\end{equation}
Remark that Eq.~(\ref{pseudo-schro-theta}) has the form of the 1D SE with a fictitious potential
\begin{equation}
V(\vartheta )={\frac{\hbar ^{2}}{2m_0}}{\frac{(m_{l}^{2}-1/4)}{\sin ^{2}\vartheta }}\ ,
\end{equation}
and an energy
\begin{equation}
E_{\vartheta }=(\lambda +{\frac{1}{4}}){\frac{\hbar ^{2}}{2m_0}}\ .
\end{equation}
Remark also that Eq. (\ref{1d-schro-phi}) has the same form as the 1D
SE with a vanishing potential and an energy
equal to $m_{l}^{2}\hbar ^{2}/2m$. \noindent Because
Eqs.~(\ref{1d-schro-radial}), (\ref{1d-schro-theta}) and
(\ref{1d-schro-phi}) come to the form of the 1D SE,
in order to obtain three separated 1D-QSHJEs, let us write the
functions $\mathcal{X}(r)$, $\mathcal{T}(\vartheta)$ and
$F(\varphi )$ as
\begin{equation}\label{xi}
\hskip-5mm\Psi(r,\vartheta,\varphi)={\cal A}(r)\xi (\vartheta )\eta (\varphi )
\left[ \alpha e^{{\frac{i}{\hbar }}[Z(r)+L(\vartheta )+M(\varphi )]}+\beta
e^{-{\frac{i}{\hbar }}[Z(r)+L(\vartheta )+M(\varphi )]}\right]\, ,
\end{equation}
where ${\cal A}(r)\xi (\vartheta )\eta (\varphi )=A(r,\vartheta,\varphi)$ is the amplitude
of the wave function, and $S_0(r,\vartheta,\varphi)=Z(r)+L(\vartheta )+M(\varphi )$
is the total reduced action. (Expressing
the total reduced action as a sum of three 1D reduced actions is argued in
Ref. \cite{Dja4} in the case of cartesian symmetry potentials. For the case of the central
potential, the separation of the variables in the reduced action still valid and
one can prove it by proceeding by the same way as in Ref. \cite{Dja4}. \noindent By replacing Eq.~(\ref{xi}) in Eqs.~(\ref{1d-schro-radial}),
(\ref{1d-schro-theta}) and (\ref{1d-schro-phi}) respectively, we
obtain exactly as we had proceeded in 1D
\begin{equation}\label{1d-qshje-radial}
{\frac{1}{2m_0}}\left( {\frac{dZ}{dr}}\right) ^{2}-{\frac{\hbar ^{2}}{4m_0}}%
\left\{ Z,r\right\} +V(r)+{\frac{ l(l+1) \hbar
^{2}}{2m_0r^{2}}}=E\ ,
\end{equation}
\begin{equation}\label{1d-qshje-theta}
\left( {\frac{dL}{d\vartheta }}\right) ^{2}-{\frac{\hbar
^{2}}{2}}\left\{ L,\vartheta \right\} +{\frac{(m_{l}^{2}-1/4)}{\sin ^{2}\vartheta }}\hbar ^{2}=\left(l(l+1)
+{\frac{1}{4}}\right)\hbar ^{2}\;,
\end{equation}
\begin{equation}\label{1d-qshje-phi}
\left( {\frac{dM}{d\varphi }}\right) ^{2}-{\frac{\hbar
^{2}}{2}}\left\{ M,\varphi \right\} =m_{l}^{2}\hbar ^{2}\ .
\end{equation}
Eqs.~(\ref{1d-qshje-radial}), (\ref{1d-qshje-theta}) and
(\ref{1d-qshje-phi}) represent the components of the QSHJE in 3-D
for a central potential. These equations contain the Schwarzian
derivatives of the functions $Z$, $L$ and $M$ with respect to $r$,
$\vartheta$ and $\varphi$ respectively. Note that, by deducing the
value of $\lambda $ from Eq.~(\ref{1d-qshje-radial}), and the
value of $m_{l}$ from Eq.~(\ref{1d-qshje-phi}), then substituting it
in Eq.~(\ref{1d-qshje-theta}), one obtains
\begin{eqnarray}\label{3d-qshje-sphere-action}
{\frac{1}{2m_0}}\left( \vec{\nabla}_{r,\vartheta ,\varphi }S_{0}\right) ^{2}-{%
\frac{\hbar ^{2}}{4m_0}}\left[ \left\{ S_{0},r\right\}
+{\frac{1}{r^{2}}}\left\{ S_{0},\vartheta \right\} +
\right.
\hskip5mm &&
\nonumber \\
\left. {\frac{1}{%
r^{2}\sin ^{2}\vartheta }}\left\{ S_{0},\varphi \right\} \right] +V(r)-{%
\frac{\hbar ^{2}}{8m_0\ r^{2}}}-{\frac{\hbar ^{2}}{8m_0\ r^{2}\sin
^{2}\vartheta }}=E\; .
\end{eqnarray}
Eq.~(\ref{3d-qshje-sphere-action}) represents the QSHJE in 3-D for
a central potential. At the classical limit $(\hbar \to 0)$,
Eq.~(\ref{3d-qshje-sphere-action}) goes to the Classical Stationary Hamilton-Jacobi Equation
(CSHJE). Remark that taking the classical limit in
Eqs.~(\ref{1d-qshje-radial}), (\ref{1d-qshje-theta}) and
(\ref{1d-qshje-phi}) makes $dL/d\vartheta $ and $dM/d\varphi $
vanish, then, one cannot obtain the CSHJE. We deduce
that the classical limit must be taken into
Eq.~(\ref{3d-qshje-sphere-action}). \noindent Note also that
separating variables in Eq.~(\ref{3d-qshje-sphere-action}), we get to
Eqs.~(\ref{1d-qshje-radial}), (\ref{1d-qshje-theta}) and
(\ref{1d-qshje-phi}) which lead to Eqs.~(\ref{1d-schro-radial}),
(\ref{1d-schro-theta}) and (\ref{1d-schro-phi}). After separating
variables in Eq.~(\ref{3d-qshje-sphere-action}), three constants
of motion appear, they are the energy $E$, $l(l+1)$ and $m_{l}$.

\vskip0.5\baselineskip The solutions of
Eqs.~(\ref{1d-qshje-radial}), (\ref{1d-qshje-theta}) and
(\ref{1d-qshje-phi}) are respectively
\begin{equation}\label{zed}
Z(r)=\hbar \arctan \left\{ \alpha_r {{\cal X}_2 \over {\cal
X}_1}+b_r\right\}= \hbar \arctan \left\{ a_r \int{dr \over {\cal
X}^2_1}+b_r\right\}\;,
\end{equation}
\begin{equation}\label{ell}
L(\vartheta )=\hbar \arctan \left\{ \alpha_\vartheta {{\cal T}_2
\over {\cal T}_1}+b_\vartheta\right\}=\hbar \arctan \left\{ a_\vartheta \int{d\vartheta
\over {\cal T}^2_1}+b_\vartheta\right\} \;,
\end{equation}
\begin{equation}\label{emm}
M(\varphi )=\hbar \arctan \left\{  \alpha_\varphi {F_2 \over
F_1}+b_\varphi \right\}=\hbar \arctan \left\{ a_\varphi \int{d\varphi \over
F^2_1}+b_\varphi\right\} \;,
\end{equation}
where $\alpha_{r}$, $b _{r}$, $\alpha _{\vartheta }$, $b
_{\vartheta }$, $\alpha_{\varphi }$, $b _{\varphi }$, $a_r$,
$a_\vartheta$ and $a_\varphi$ are real
constants. ${\cal X}_{1}$ and ${\cal X}_{2}$ are two real
independent solutions of Eq.~(\ref{pseudo-schro-rad}). ${\cal
T}_{1}$ and ${\cal T}_{2}$ are two real independent solutions of
Eq.~(\ref{pseudo-schro-theta}). $F_1$ and $F_2$ are two real
independent solutions of Eq.~(\ref{1d-schro-phi}).

Before introducing the quantum law of motion, we would like to
remind the fact that for the central potential, only the radial
function ${\cal X}$ depends on the form of the potential,
meaning that the two angular functions ${\cal T}$ and $F$ can
be established without considering the form of the central
potential. Indeed, if we take a look on the table of solutions of
the SE for such a potential we should write
the angular function ${\cal T}_1$ as \cite{Nik}
\begin{equation}\label{thetawave}
{\cal T}^{(lm_l)}_1(\vartheta)=(1-\cos^2(\vartheta))^{{|m_l| \over
2}+{1 \over 4}}\; {d^{l+|m_l|}\ \ \ \ \ \ \ \ \ \over
d(\cos(\vartheta))^{l+|m_l|}}\left[(1-\cos^2(\vartheta))^l\right]\; ,
\end{equation}
and the angular function $F_1$ as
\begin{eqnarray}\label{phiwave}
F_1(\varphi)=\sin(|m_l|\varphi) \hskip15mm if \hskip5mmm_l\neq 0
&& \nonumber \\   = 1 \hskip27mm if \hskip5mmm_l= 0\; ,
\end{eqnarray}
up to normalization constants.
\vskip\baselineskip \noindent {\bf 3.\ \ The Quantum law of motion
for a particle moving under the central potential}
\vskip0.5\baselineskip
Now let us investigate the quantum law of motion for the central potential case. Knowing that the
3D quantum law of motion is given by Eq.~(\ref{3d-qlmotion})
and using the spherical coordinates in it, we may write
\begin{equation}\label{3d-qlmotion-sphere}
\dot{r}{dZ \over dr}+\dot{\vartheta}{dL \over
d\vartheta}+\dot{\varphi}{dM \over d\varphi}=2\left[ E-V(r)\, .
\right]
\end{equation}
As in classic mechanics, we do not use directly
Eq.~(\ref{3d-qlmotion-sphere}) to plot the trajectories. We must
first, separate the variables $r$, $\vartheta$ and $\varphi$. This
seems not possible from Eq.~(\ref{3d-qlmotion-sphere}), and must
be done in the 3D QSHJE (Eq.~(\ref{3d-qshje-sphere-action})). We
noted in Sec. 2 that the separation of variables in
Eq.~(\ref{3d-qshje-sphere-action}) leads directly to
Eqs.~(\ref{1d-qshje-radial}), (\ref{1d-qshje-theta}) and
(\ref{1d-qshje-phi}). In Ref. \cite{Dja1}, we have considered a
deformation of the space like for general relativity.
This idea was based on the quantum coordinate introduced in a
earlier work of Faraggi and Matone \cite{FM1}. We treated
the 3D quantum systems with a general potential by using the
mechanics of a deformed space. This allowed us to introduce a
quantum Lagrangian and derive the quantum law of motion
(Eq.~(\ref{3d-qlmotion})). For the central potential, we proceed
with the same manner. Thus, let us write
Eq.~(\ref{3d-qshje-sphere-action}) as
\begin{eqnarray}\label{twrdlaw}
\left({dZ \over dr}\right)^{2}\;  g_{rr}(r)\, +\; {1 \over r^2}\,
\left({dL \over d\vartheta}\right)^{2} g_{\vartheta\vartheta}(\vartheta) \, \, +
\hskip40mm\nonumber\\
\, {1 \over r^2\, sin^2(\vartheta)}\, \left({dM \over
d\varphi}\right)^{2}\;  g_{\varphi\varphi}(\varphi)\, =
2m_0[E-V(r)] \; ,
\end{eqnarray}
where
$$ g_{rr}(r)=1-{\hbar^2 \over 2}\left({dZ \over
dr}\right)^{-2}\{Z,r\}\; , \hskip15mm $$
\begin{equation}\label{tensor}
g_{\vartheta\vartheta}(\vartheta)=1-{\hbar^2 \over
2}\left({dL \over d\vartheta}\right)^{-2}\left(\{L,\vartheta\}-{1
\over  2}\right)\; , \hskip5mm
\end{equation}
and
$$ g_{\varphi\varphi}(\varphi)=1-{\hbar^2 \over 2}\left({dM
\over d\varphi}\right)^{-2}\left(\{M,\varphi\}-{1 \over
2}\right)\; , $$
are real functions depending on $r$, $\vartheta$ and $\varphi$
respectively. They represents the non vanishing components of the
metric tensor of the curved space created by the central
potential. Now, if one identifies the energy $E$ to the quantum
hamiltonian $H$, and using the canonical equations $\partial H/\partial P_r=\dot{r}$,
$\partial H/\partial P_{\vartheta}=\dot{\vartheta}$ and $\partial H/\partial P_{\varphi}=\dot{\varphi}$
he gets
\begin{equation}\label{radial_tensor}
{dZ\over dr} =m_0\dot{r}g_{rr}\; ,
\end{equation}
\begin{equation}\label{theta_tensor}
{dL \over d\vartheta }=
m_0r^2{\dot{\vartheta}}g_{\vartheta\vartheta}\; ,
\end{equation}
\begin{equation}\label{phi_tensor}
{dL \over d\varphi }=
m_0r^2\sin^2(\vartheta){\dot{\varphi}}g_{\varphi\varphi}\; .
\end{equation}
Taking expressions of $g_{rr}$, $g_{\vartheta\vartheta}$ and $g_{\varphi\varphi}$ from the last three equations and replacing them into Eq.~(\ref{twrdlaw}), we deduce Eq.~(\ref{3d-qlmotion-sphere}), while replacing them into into Eqs.~(\ref{1d-qshje-radial}),
(\ref{1d-qshje-theta}) and (\ref{1d-qshje-phi}) respectively after
taking into account of relations (\ref{tensor}), we find
\begin{equation}\label{radial-qlmotion}
{dZ \over dr}\; \dot{r}=2\left[E-V(r)-{l(l+1)\hbar^2\over
2m_0r^2}\right]\; ,
\end{equation}
\begin{equation}\label{theta-qlmotion}
{dL \over d\vartheta}\; \dot{\vartheta}={l(l+1)\hbar^2 \over
m_0r^2}-{(m^2_l-{1/4})\hbar^2 \over m_0r^2\sin^2(\vartheta)}\; ,
\end{equation}
\begin{equation}\label{phi-qlmotion}
{dM \over d\varphi}\; \dot{\varphi}={(m^2_l-{1/4})\hbar^2 \over
m_0r^2\sin^2(\vartheta)}\; .  \hskip18mm
\end{equation}
The three above relations represent the decomposition of
Eq.~(\ref{3d-qlmotion-sphere}) with respect to the variables $r$,
$\vartheta$ and $\varphi$. Indeed, if one takes the sum of
Eqs.~(\ref{radial-qlmotion}), (\ref{theta-qlmotion}) and
(\ref{phi-qlmotion}) he finds Eq.~(\ref{3d-qlmotion-sphere}). In
fact, these equations describe the dynamical behaviour of a
particle moving under the central potential, each one describes it
in the corresponding coordinate ($r$, $\vartheta$ or $\varphi$).
\noindent Note that only Eq.~(\ref{radial-qlmotion}) depends on
the nature of the central potential $V(r)$ while
Eqs.~(\ref{theta-qlmotion}) and (\ref{phi-qlmotion}) do not depend
on it. However, they depend implicitly on the radial potential via the radial position $r$.

As an application of the 3D quantum law of motion
(Eqs.~(\ref{3d-qlmotion-sphere}), ~(\ref{radial-qlmotion}),
(\ref{theta-qlmotion}) and (\ref{phi-qlmotion})), we propose in
the following sections to plot the trajectories of an electron
moving under the action of the Hydrogen potential (Hydrogen atom).
This will allow us to acquire more comprehension of the 3D
quantum law of motion.
\vskip\baselineskip \noindent {\bf 4.\ \ Classical trajectories
of a particle moving under a central potential}
\vskip0.5\baselineskip
Before plotting the QTs, it will be useful to
present the classical Hamilton-Jacobi formulation of mechanics and
also to plot the classical trajectory of a particle moving under
the central potential. The Hamilton-Jacobi formulation is based on
the CHJE. In the case of a central potential,
the separation of the variables in the CHJE gives
\begin{equation}\label{clas_mom-radial}
\left({dZ \over dr}\right)^2=2m_0\left[E-V(r)-{\alpha \over
2m_0r^2}\right]\; ,
\end{equation}
\begin{equation}\label{clas_mom-theta}
\left({dL \over d\vartheta}\right)^2={\alpha }-{\beta^2 \over
\sin^2(\vartheta)}\; , \hskip20mm
\end{equation}
and
\begin{equation}\label{clas_mom-phi}
\left({dM \over d\varphi}\right)^2={\beta^2}\; . \hskip25mm
\end{equation}
\noindent $\alpha$ and $\beta$ are real constants. As we know,
the momenta $dZ/dr$, $dL/d\vartheta$ and $dM/d\varphi$ are
connected to the speeds $\dot{r}$, $\dot{\vartheta}$ and
$\dot{\varphi}$ as follows
\begin{equation}
{dZ \over dr}=m_0\dot{r}\; , \hskip16mm
\end{equation}
\begin{equation}
{dL \over d\vartheta}=m_0r^2\dot{\vartheta}\; ,\hskip12mm
\end{equation}
\begin{equation}
{dM \over d\varphi}=m_0r^2\sin^2(\vartheta)\dot{\varphi}\; .
\end{equation}
Taking these last equations into Eqs.~(\ref{clas_mom-radial}),
(\ref{clas_mom-theta}) and (\ref{clas_mom-phi}), we find
\begin{equation}\label{rdot}
{\dot{r}}=\pm\sqrt{2 \over m_0}\sqrt{E-V(r)-{\alpha \over
2m_0r^2}}\; ,\hskip0mm
\end{equation}
\begin{equation}\label{thetadot}
{\dot{\vartheta}}=\pm{1 \over m_0r^2}\sqrt{{\alpha }-{\beta^2
\over \sin^2(\vartheta)}}\; ,\hskip10mm
\end{equation}
\begin{equation}\label{phidot}
{\dot{\varphi}}=\pm{\beta \over m_0r^2\sin^2(\vartheta)}\;
,\hskip20mm
\end{equation}
The $\pm$ signs in Eqs.~(\ref{rdot}), (\ref{thetadot}) and
(\ref{phidot}) indicate that the motion of the particle may be
in either direction on the $r$ axis, $\vartheta$ circle and $\varphi$
semicircle respectively. Now, we can integrate Eqs.~(\ref{rdot}),
(\ref{thetadot}) and (\ref{phidot}) to deduce the three time
equation with respect to the coordinates $r$, $\vartheta$ and
$\varphi$, but before, we write these equations as
\begin{equation}\label{rtime}
\sqrt{2 \over m_0}\; {dt}=\pm{ dr \over \sqrt{E-V(r)-{\alpha /
(2m_0r^2)}}}\; ,\hskip0mm
\end{equation}
\begin{equation}\label{thetatime}
 {dt \over m_0r^2}=\pm {d\vartheta \over \sqrt{{\alpha }-{\beta^2
 /\sin^2(\vartheta)}}}\; ,\hskip10mm
\end{equation}
\begin{equation}\label{phitime}
 {dt \over m_0r^2\sin^2(\vartheta)}=\pm {d\varphi \over \beta}\; .\hskip20mm
\end{equation}
\begin{figure}
\def\put(#1,#2)#3{\leavevmode\rlap{\hskip#1\unitlength\raise#2\unitlength\hbox{#3}}}
\centerline{\includegraphics[height=7cm,width=7cm,angle=270]{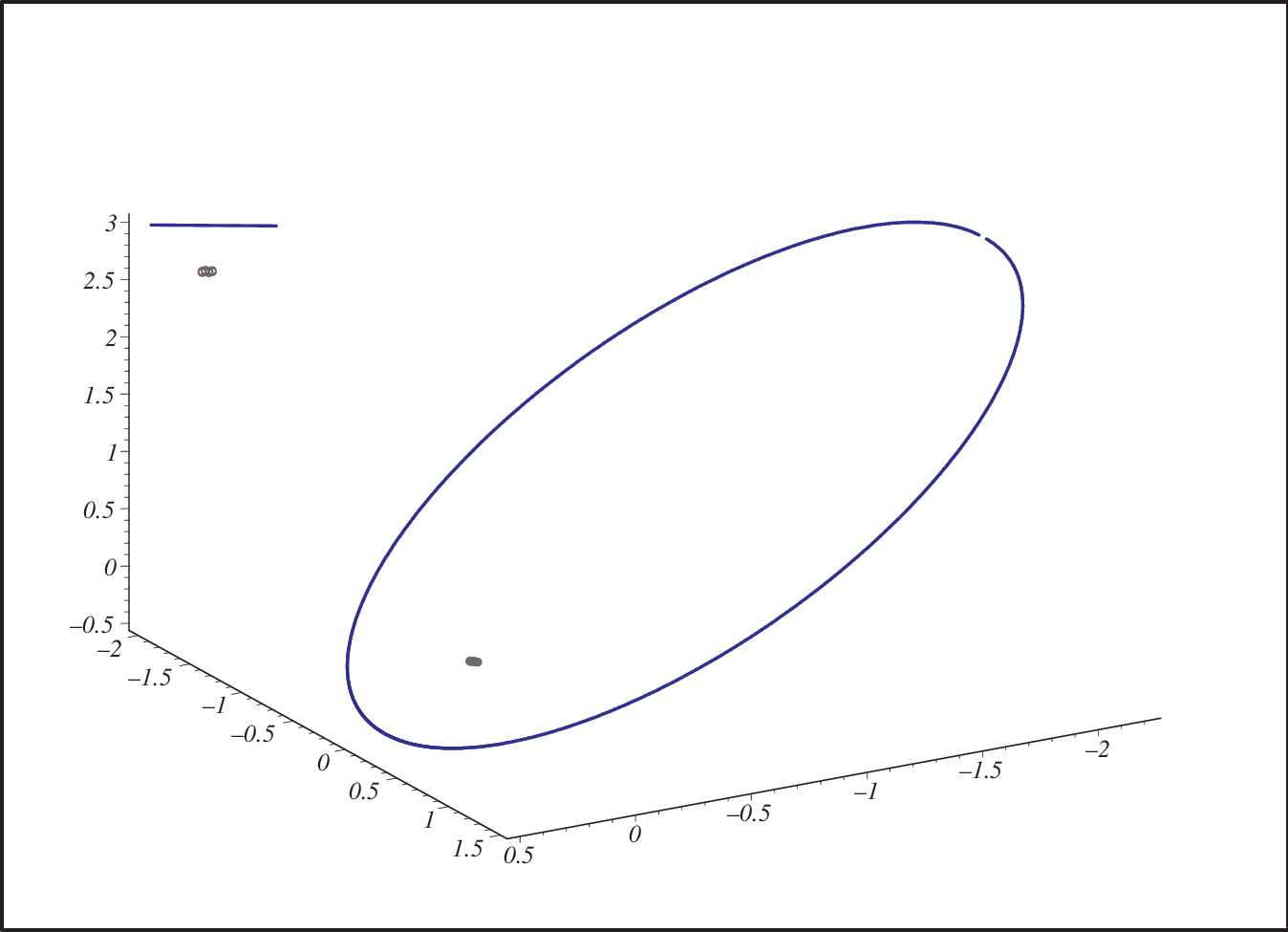}
\put(-192,-42.){\tiny{$z$($\times 10^{-10}$ m)}}
\put(-140,-188){\tiny{$y$($\times 10^{-10}$ m)}}
\put(-37.,-160){\tiny{$x$($\times 10^{-10}$ m)}}
\put(-170,-49){\tiny{ Classical trajectory}}
\put(-170,-59){\tiny{ Nucleus position }}}
\raisebox{-1.6cm}{\centerline{\vbox{ \hsize=7cm\noindent\scriptsize   Fig1. Classical 3D trajectory of the Hydrogen's
                electron. In order to consider the atomic scalars,
                we have chosen in Eqs (71-73), $E=-13.6/4$\, eV,
                $\alpha=2\, \hbar^2$ and $\beta=(\sqrt{3}/2)\, \hbar^2$.
                It is an ellipse, and we see clearly that the motion is
                done in a plan.}}}\label{class1}
\end{figure}

We remark that Eq.~(\ref{rtime}) can be integrated directly after
knowing the form of the central potential (either analytically or
numerically). However, to integrate Eq.~(\ref{thetatime}), we must
first write $r$ -already integrated from Eq.~(\ref{rtime})- in
function of $t$. Also, to integrate Eq.~(\ref{phitime}),
we should set the variables $r$ and $\vartheta$ in function of
$t$. In spite of the importance of the time equations, we will
focus on the spatial trajectory equations. These last's can be
derived directly from Eqs.~(\ref{rtime}), (\ref{thetatime}) and
(\ref{phitime}) after parameterizing the time $t$. Thus, we deduce
\begin{equation}\label{clasrdyn}
\sin^2(\vartheta)d\varphi=\pm{\beta dr \over
r^2\sqrt{2m_0(E-V(r))-\alpha/r^2}}\; ,
\end{equation}
\begin{equation}\label{clasthetadyn}
{d\vartheta \over \sqrt{{\alpha }-{\beta^2
 /\sin^2(\vartheta)}}}=\pm{ dr \over r^2\sqrt{2m_0(E-V(r))-{\alpha / r^2}}}\; ,\hskip0mm
\end{equation}
\begin{equation}\label{clasphidyn}
d\varphi=\pm{ d\vartheta \over
\sin(\vartheta)\sqrt{(\alpha/\beta^2)\sin^2(\vartheta)-1}}\; .
\hskip10mm
\end{equation}
Integrating these last equations we find the trajectory equation
${\cal F}(r, \vartheta, \varphi)=0$.
\noindent In order to compare the classical trajectories with the
QTs, we plot in Fig.1 the trajectory of an
electron under the atomic potential (Hydrogen atom)
\begin{equation}\label{hydpot}
V(r)=-{k\; e^2 \over r}\; ,
\end{equation}
where $k=9 . 10^9 S.I$ is the electric constant and $e$ the charge
of the electron. We give the energy $E$ the value -(13.6/4) eV,
which corresponds in the quantum case to the energy of the first
excited state ($n=2$, $n$ being the principle
number).This is in order to compare the
classical trajectory (Fig.1) with the QTs
corresponding to this value of energy. For the same reason, we
have chosen the constants $\alpha=2\, \hbar^2$ and
$\beta=(\sqrt{3}/2)\, \hbar$ equal to the values of $l(l+1)\,
\hbar^2$ and $\sqrt{m^2_l-1/4}\; \hbar$ respectively, for $l=1$
and $m_l=1$. Then, the QTs corresponding to the
classical trajectory are those plotted for $n=2$, $l=1$ and
$m_l=1$ (Figs.7 and 8).

\noindent The classical trajectory
shown in Fig.1 is an inclined ellipse. This trajectory is
obviously famous since it is analogous to the trajectory of a planet submitted
to the gravitation force of the sun.
\vskip\baselineskip \noindent {\bf 5.\ \ Quantum orbits of the
Hydrogen's electron } \vskip0.5\baselineskip
In this section we will plot the 3D QTs. First,
let us rewrite Eqs.~(\ref{radial-qlmotion}),
(\ref{theta-qlmotion}) and (\ref{phi-qlmotion}) as follows
\begin{equation}\label{quartime}
2\; dt=\left[E-V(r)-{l(l+1)\hbar^2 \over 2m_0r^2}\right]^{-1}{dZ
\over dr}\; dr\; ,\hskip5mm
\end{equation}
\begin{equation}\label{quathetatime}
dt={m_0r^2 \over \hbar^2}\left[{l(l+1)}-{(m^2_l-{1/4}) \over
\sin^2(\vartheta)}\right]^{-1}{dL \over d\vartheta}\; d\vartheta\;
,
\end{equation}
\begin{equation}\label{quaphitime}
dt={m_0r^2\sin^2(\vartheta) \over (m^2_l-{1/4})\hbar^2}{dM \over
d\varphi}\; d\varphi\; . \hskip26mm
\end{equation}
By integrating Eqs.~(\ref{quartime}), (\ref{quathetatime}) and
(\ref{quaphitime}) we find the time equations. In order to plot
the QTs, we need to find the trajectories
equations. In this order, we parameterize the time $dt$ into
Eqs.~(\ref{quartime}), (\ref{quathetatime}) and (\ref{quaphitime})
so as to obtain
\begin{equation}\label{quardyn}
{\sin^2(\vartheta) \over (m^2_l-{1/4})}{dM \over d\varphi}\;
d\varphi=\left[{2m_0 \over \hbar^2}\; r^2 (E-V(r)) -l(l+1)
\right]^{-1}{dZ \over dr}\; dr\; ,\hskip5mm
\end{equation}
\begin{eqnarray}\label{quathetadyn}
\left[{l(l+1)}-{(m^2_l-{1/4}) \over
\sin^2(\vartheta)}\right]^{-1}{dL \over d\vartheta}\;
d\vartheta=\hskip40mm\nonumber\\
\left[{2m_0 \over \hbar^2}\; r^2 (E-V(r)) -l(l+1) \right]^{-1}{dZ
\over dr}\; dr\; ,
\end{eqnarray}
\begin{equation}\label{quaphidyn}
{dM \over d\varphi}\; d\varphi=\left[{l(l+1)\over (m^2_l-{1/4})}\;
\sin^2(\vartheta)-1\right]^{-1}{dL \over d\vartheta}\;
d\vartheta\; .\hskip25mm
\end{equation}
In what follows Eqs.~(\ref{quardyn}), (\ref{quathetadyn}) and
(\ref{quaphidyn}) are used to plot the 3D quantum spatial
trajectories (orbits). Expressions of the quantum momenta ${dZ /
dr}$, ${dL / d\vartheta}$ and ${dM / d\varphi}$ are deduced
directly from the corresponding quantum reduced actions $Z$, $L$
and $M$ (Eqs.~(\ref{zed}), (\ref{ell}) and (\ref{emm})
respectively)
\begin{equation}\label{radmom}
{dZ \over dr}=\pm{\hbar a_r {\cal X}_1^{-2}\over \left(a_r\int {dr \over
{\cal X}^2_1} +b_r\right)^2+1}\; ,
\end{equation}
\begin{equation}\label{thetamom}
{dL \over d\vartheta}=\pm{\hbar a_\vartheta {\cal T}_1^{-2}\over
\left(a_\vartheta\int {d\vartheta \over {\cal T}^2_1}
+b_\vartheta\right)^2+1}\; ,
\end{equation}
\begin{equation}\label{phimom}
{dM \over d\varphi}=\pm{\hbar a_\varphi F_1^{-2} \over \left(a_\varphi\int
{d\varphi \over F^2_1} +b_\varphi \right)^2+1}\ .
\end{equation}
Knowing the expressions of the functions ${\cal X}_1$, ${\cal
T}_1$ and $F_1$, the system
of the three differential equations (\ref{quardyn}-\ref{quaphidyn}) can be solved
numerically for the different values of the constants $a_r$,
$b_r$, $a_\vartheta$, $b_\vartheta$, $a_\varphi$ and $b_\varphi$.
The $\pm$ sign in the expressions of the quantum momenta shows the
fact that the motion of the particle may be in either direction of
the $r$ axis, $\vartheta$ semicircle and $\varphi$ circle.

\noindent {\bf Note:} In fact two of Eqs.~(\ref{quardyn}),
(\ref{quathetadyn}) and (\ref{quaphidyn}) are sufficient to
integrate and determine the quantum spatial trajectories since the
third equation is automatically satisfied when the two others are.
\vskip0.5\baselineskip Now, let us consider the quantum motion of
the Hydrogen's electron. We take up the expression of the atomic
potential given in Eq.~(\ref{hydpot}). In this case, the solving
of Eq.~(\ref{pseudo-schro-rad}) is obvious and gives for the
radial function ${\cal X}$ the following expression \cite {Nik}
\begin{equation}\label{radwave}
{\cal X}^{(nl)}_1(r)=\exp\left({r \over na_{o}}\right)\; \;
\left({r \over a_o }\right)^{-l}\; {d^{n-l-1} \over
dr^{n-l-1}}\left[\left({r \over a_o }\right)^{n+l}\exp\left(-{2r \over
na_{o}}\right)\right]
\end{equation}
$n$ being an integer number representing the principle quantum
number, and
$$
a_{o}={\ \hbar^2 \over m_0 k e^2}
$$
is the Bohr's radius. The function ${\cal X}^{(nl)}_1$ corresponds
to the energy
$$
E_n=-{k\, e^2 \over 2\, n^2\, a_{o}}\; .
$$
Fixing the values of $n$, $l$ and $m_l$ into Eqs. (\ref{radwave}), (\ref{thetawave}) and
(\ref{phiwave}) and  substituting these equations into Eqs. (\ref{radmom}), (\ref{thetamom})
and (\ref{phimom}) respectively, then replacing these lasts'  into Eqs. (\ref{quardyn}-\ref{quaphidyn})
after taking account of the above expressions of the Bohr's radius and the energy, we get a system of three differential equations which after integrating leads to the 3D QTs
of the electron for the corresponding bound
state $(n,l,m_l)$. Naturally, these trajectories depend on the
constants $a_r$, $b_r$, $a_\vartheta$, $b_\vartheta$, $a_\varphi$
and $b_\varphi$. So, for several sets of those constants it
corresponds many trajectories (See Figs. (2-7)). In other words,
for each bound state many trajectories exist exactly as we have
found for the 1D quantum motions. This shows the existence of microstates that
the QSHJE detects and that are not detected by the SE as it is shown in 1D in Ref. \cite{Floyd}.
We will show this by plotting
the trajectories of three bound states, the fundamental state
$(n=1,\; l=0,\; m_l=0)$, and the excited bound states $(n=2,\;
l=0,\; m_l=0)$ and ($n=2,\; l=1,\;
m_l=1)$\footnote{The considered excited states in this article do
not obeys to the selection law that we consider in spectroscopy of
the Hydrogen atom. We just use these states in order to make a
simple description of the QTs for different
values of the quantum numbers $n$, $l$ and $m_l$.}. With this aim,
for each state we proceed as follows:

\noindent First, we explicit
the functions ${\cal X}^{(nl)}_1$, ${\cal T}^{(lm)}_1$ and $F_1^{(m_l)}$
from Eqs. (\ref{radwave}), (\ref{thetawave}) and
(\ref{phiwave}). Secondly, we integrate numerically
Eq.~(\ref{quartime}) and plot the time equation $r(t)$ for
different values of the constants $a_r$ and $b_r$. After,
by numerical methods we integrate the two Eqs. (\ref{quardyn}) and (\ref{quaphidyn}) for
different values of the constants by setting $\varphi$ as the free
variable varying from $-\pi$ to $\pi$ or multiple values of this angle. Then, we plot the resulted data to show the quantum spatial
trajectories.
\newpage
\vskip0.5\baselineskip \noindent {\bf a- the fundamental state}
\vskip0.3\baselineskip
%
%
\begin{figure}\label{time100}
\def\put(#1,#2)#3{\leavevmode\rlap{\hskip#1\unitlength\raise#2\unitlength\hbox{#3}}}
\centerline{\includegraphics[height=7cm,width=7cm,angle=270]{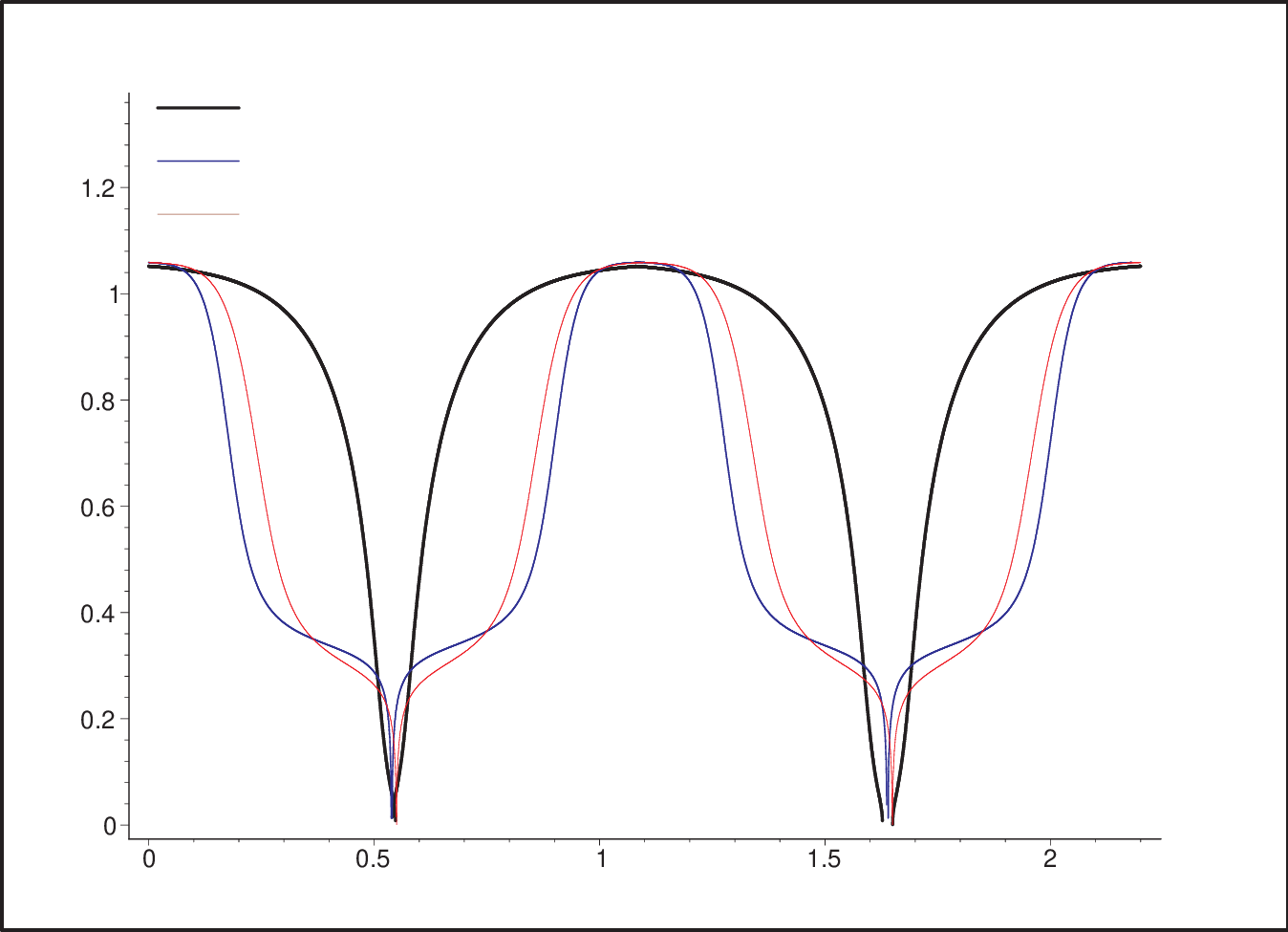}
\put(-35,-188){\tiny{$t$($\times 10^{-16}$ s)}}
\put(-200.,-15){\tiny{$r$($\times 10^{-10}$ m)}}
\put(-170,-23) {\tiny $a_r=0.18,b_r=0.85$}
\put(-174,-34) {\tiny $a_r=3,b_r=0.5$}
\put(-177,-44) {\tiny $a_r=2,b_r=1$}
}
\raisebox{-1.5cm}{\centerline{\vbox{ \hsize=8cm\noindent\scriptsize   Fig2. Quantum radial time trajectories for the
                Hydrogen's electron in the fundamental state
                ($E=-13.6$\, eV,$n=1$, $l=0$, $m_l=0$). For all
                the curves, we have chosen $r(t=0)= 2a_o$ ($a_o=0.52917\,
                .\, 10^{-10}$ m  is the Bohr radius).}}}
\end{figure}
%
%

For this state, we have $n=1$, $l=0$ and $m_l=0$. The functions
${\cal X}^{(10)}_1$, ${\cal T}^{(00)}_1$ and $F_1^{(0)}$ are deduced
from Eqs. (\ref{radwave}), (\ref{thetawave}) and
(\ref{phiwave}) as
\begin{equation}\label{r10}
{\cal X}^{(10)}_1(r)={r \over a_o }\; \exp\left(-{r \over
a_{o}}\right)\; ,
\end{equation}
\begin{equation}\label{theta10}
{\cal T}^{(00)}_1(\vartheta)=\sin^{1 \over 2}(\vartheta)\;
,\hskip7mm
\end{equation}
\begin{equation}\label{phi10}
F^{(0)}_1(\varphi)=1\; .\hskip18mm
\end{equation}
Replacing Eqs. (\ref{r10}), (\ref{theta10}) and (\ref{phi10})
into Eqs. (\ref{radmom}), (\ref{thetamom}) and (\ref{phimom}), we find
\begin{equation}\label{radmom10}
{dZ \over dr}=\pm{\hbar a_r\; \left({r \over a_o
}\right)^{-2}e^{2r \over a_o } \over (a_r I_1(r)+b_r)^2+1}\; ,
\end{equation}
\begin{equation}\label{thetamom10}
{dL \over d\vartheta}=\pm{\hbar a_\vartheta \sin^{-1}(\vartheta)
\over [a_\vartheta \log(|\tan({\vartheta \over
2})|)+b_\vartheta]^2+1}\; ,
\end{equation}
\begin{equation}\label{phimom10}
{dM \over d\varphi}=\pm{\hbar a_\varphi \over (a_\varphi \varphi
+b_\varphi)^2+1}\ , \hskip13mm
\end{equation}
where
$$
I_1(r)=\int \left({a_o \over r}\right)^2 \exp\left({2r \over a_o }\right)\, dr=-{a_o^2\over r}\exp\left({2r \over a_o }\right)-2a_o \mathrm{Ei}\left(1, -2{r \over a_o }\right)\, ,
$$
$\mathrm{Ei}$ being the exponential integral function. We should notice here that we are
interested only on the real part of the function $\mathrm{Ei}$ and not on the imaginary
part, in fact, we should write ${\Re}(\mathrm{Ei})$.
\begin{figure}\label{quan100}
\def\put(#1,#2)#3{\leavevmode\rlap{\hskip#1\unitlength\raise#2\unitlength\hbox{#3}}}
\centerline{\includegraphics[height=7cm,width=8cm,angle=270]{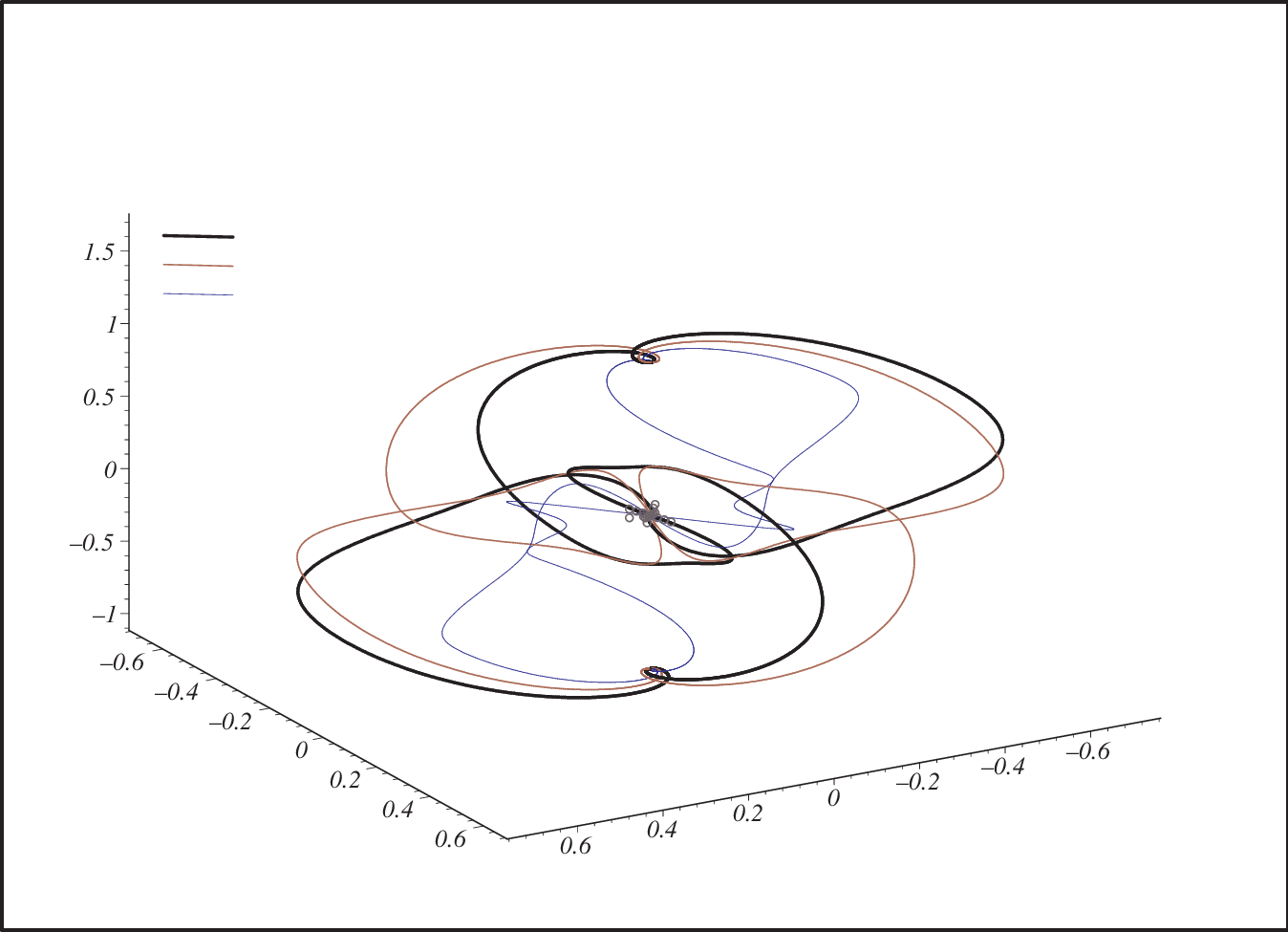}
\put(-192,-50.){\tiny{$z$($\times 10^{-10}$ m)}}
\put(-140,-214){\tiny{$y$($\times 10^{-10}$ m)}}
\put(-37.,-186){\tiny{$x$($\times 10^{-10}$ m)}}
\put(-173,-56) {\tiny $a_r=1,b_r=0,a_{\vartheta}=1,b_{\vartheta}=0,a_{\varphi}=1,b_{\varphi}=0$}
\put(-176,-65) {\tiny $a_r=2,b_r=1,a_{\vartheta}=2.4,b_{\vartheta}=-0.8,a_{\varphi}=3,b_{\varphi}=-1$}
\put(-180,-73) {\tiny $a_r=4.5,b_r=-0.6,a_{\vartheta}=1.4,b_{\vartheta}=1,a_{\varphi}=2,b_{\varphi}=0.5$}
}
\raisebox{-1.25cm}{\centerline{\vbox{ \hsize=7cm\noindent\scriptsize   Fig3.
Quantum 3D spatial trajectories of the Hydrogen's
                electron in the fundamental state ($E=-13.6$\, eV, $n=1$,
                $l=0$, $m_l=0$).}}}
\end{figure}
Replacing Eq. (\ref{radmom10}) into Eq.~(\ref{quartime}), we find
\begin{equation}\label{finalradtime}
{\hbar \over m_0}\; dt=\pm\, \left(-{r^2 \over a^2_o}+{2\, r \over
a_o}\right)^{-1}\, \left[{a_r\; a^2_oe^{2r \over a_o } \over
\left(a_r I_1(r)+b_r\right)^2+1}\right]\; dr\; ,\hskip5mm
\end{equation}
which after integrating numerically, we plot the radial time
trajectories given in Fig.2. In this figure, the electron moves
between two radial extremities $r=0$ and $r=2\, a_o$, and
describes different trajectories for the same state. All these
trajectories pass throw some points constituting nodes with the
same manner as for the 1D motions \cite{Dja2,Dja3}. The radial
position $r=0$ of the nodes correspond to the zero of the radial
function ${\cal X}^{(10)}_1$. In particular, the nodes situated
at $r=2\, a_o$ correspond to the vanishing value of the radial
velocity given by Eq. (\ref{finalradtime}). It appears that the
radial velocities at $r=0$ take not vanishing values\footnote{
In fact, if one computes the limit of the radial velocities at
$r=0$ from Eq. (\ref{finalradtime}), he will find $\infty$.
This infinite velocities appears in our equations for this
case and the next case because we did not consider relativistic effects yet.
We think that for the quantum relativistic theory corresponding to our approach
we will avoid these infinite velocities.}
meaning that, in the space, the electron when it reaches $r=0$
continues its spatial trajectory to reach the $r=2ao$ in the opposite side
of the previous position $r=2ao$. When the electron reaches this
extremity $r=2ao$, its velocity vanishes and turns back to the position
$r=0$ from which it will continue to the other extremity $r=2ao$. This
makes the number of extremities $r=2ao$ two and
the number of the spatial trajectory's cut branches four (Fig.3).
Remark that all the trajectories
oscillate about the trajectory for which the radial hidden
variables take the values $a_r=0.18$ and $b_r=0.85$.
\begin{figure}\label{time200}
\def\put(#1,#2)#3{\leavevmode\rlap{\hskip#1\unitlength\raise#2\unitlength\hbox{#3}}}
\centerline{\includegraphics[height=7cm,width=7cm,angle=270]{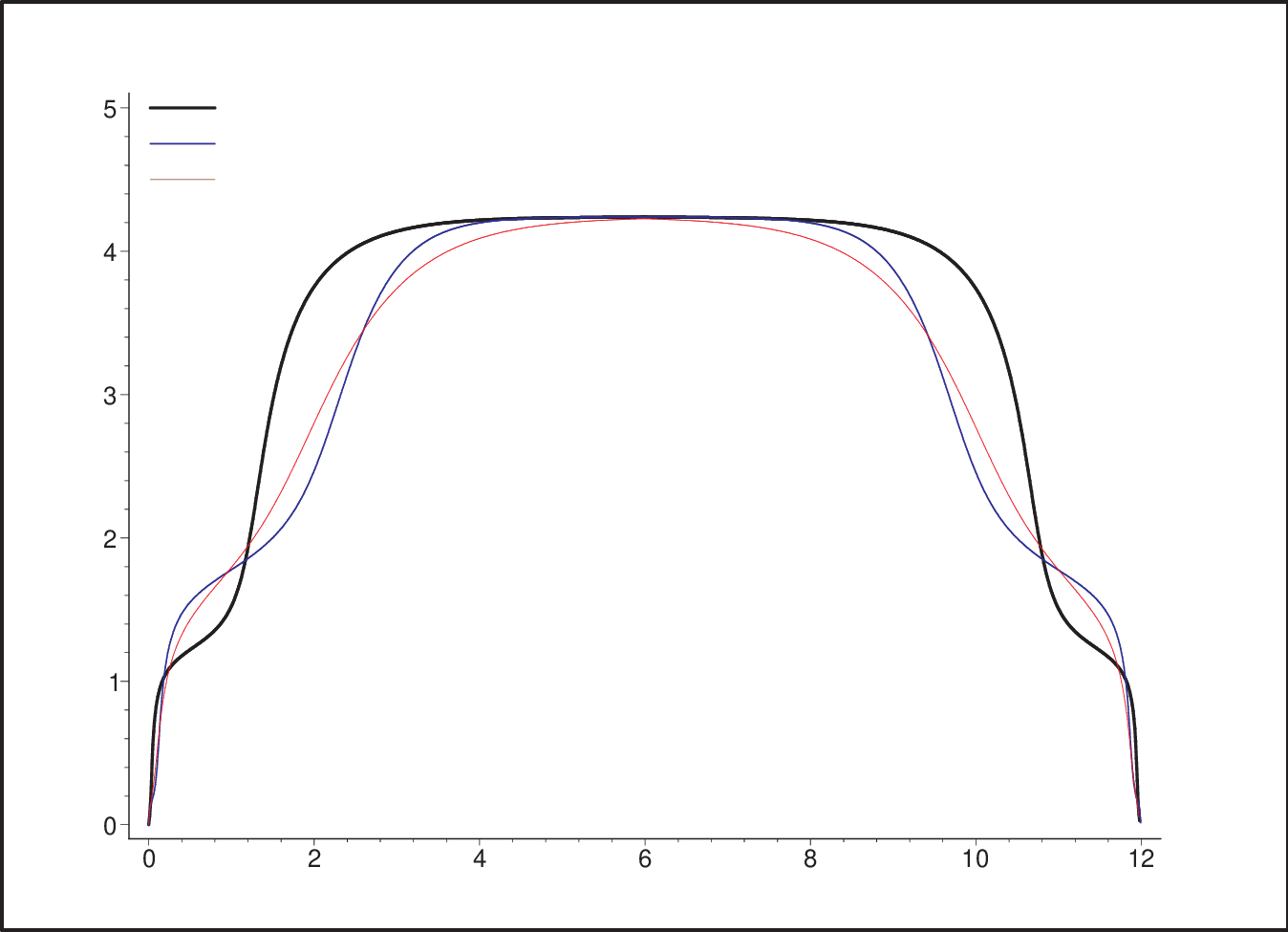}
\put(-35,-188){\tiny{$t$($\times 10^{-16}$ s)}}
\put(-200.,-15){\tiny{$r$($\times 10^{-10}$ m)}}
\put(-173,-23) {\tiny $a_r=0.6,b_r=1$}
\put(-177,-31) {\tiny $a_r=2.75,b_r=1.5$}
\put(-180,-39) {\tiny $a_r=1.5,b_r=0.75$}
}
\raisebox{-1.25cm}{\centerline{\vbox{ \hsize=8cm\noindent\scriptsize   Fig4. Quantum radial time trajectories for the Hydrogen's
                electron in the first excited state ($E=-13.6/4$\,
                eV, $n=2$,$l=0$, $m_l=0$). For all the curves,
                we have chosen $r(t=0)= 0$.}}}
\end{figure}
\noindent Now, taking Eqs. (\ref{radmom10}), (\ref{thetamom10}) and (\ref{phimom10}) into Eqs. (\ref{quardyn}) and (\ref{quaphidyn}), we get
\begin{equation}\label{raddyn10}
{4\, a_\varphi\, \sin^2(\vartheta)\,  \over (a_\varphi\, \varphi
+b_\varphi )^2+1}\, d\varphi=\pm{a_r\, a^2_o\left(-{r^2 \over
a^2_o}+{2 r\over a_o\, }\right)^{-1}\, r^{-2}\, e^{2r \over a_o}  \over
\left(a_r I_1(r)+b_r\right)^2+1}\, dr\; ,
\end{equation}
\begin{equation}\label{thetadyn10}
{a_\varphi \over (a_\varphi \varphi +b_\varphi)^2+1}\, d\varphi
=\pm{a_\vartheta \sin^{-1}(\vartheta) \over [a_\vartheta
\log(|\tan({\vartheta \over 2})|)+b_\vartheta]^2+1}\, d\vartheta
\; .
\end{equation}
Integrating the last two equations, we plot the 3D quantum spatial
trajectories given in Fig.3. This figure shows that the Hydrogen's
electron traces different trajectories which all pass from the
position $r=0$ and the two extremities $r=2ao$, these three positions
constitute nodes for the trajectories. The position $r=0$ can be
seen as the position of the nucleus in a first approximation
and we may deduce that the electron fall on the nucleus.
This is not the case because both electron and nucleus make a
relative motion compared to their gravity center, so that
the nucleus oscillate between the two radial positions
$r_{nuc}=0$ and $r_{nuc}^{max}={m_o \over m_o+M_{nuc}}\, r_{el}^{max}$,
(for the hydrogen $r_{nuc}\simeq 10^{-13}\times m$). Then the electron
has a real possibility to pass through the position $r_{el}=0$ without
collision with the nucleus. It is useful to indicate that the motion
of the electron in a state for which $m_l=0$ is a purely quantum
motion and cannot have a classical correspondent. The reason is because
in Eqs. (\ref{clas_mom-theta}) and (\ref{clas_mom-phi}), $\beta^2$ can
never take negative values while in Eq. (\ref{theta-qlmotion}) and
(\ref{phi-qlmotion}), $(m_l^2-1/4)\hbar^2$ takes a negative value for $m_l=0$.
%
%
%
%
\begin{figure}\label{quan200}
\def\put(#1,#2)#3{\leavevmode\rlap{\hskip#1\unitlength\raise#2\unitlength\hbox{#3}}}
\centerline{\includegraphics[height=7cm,width=8cm,angle=270]{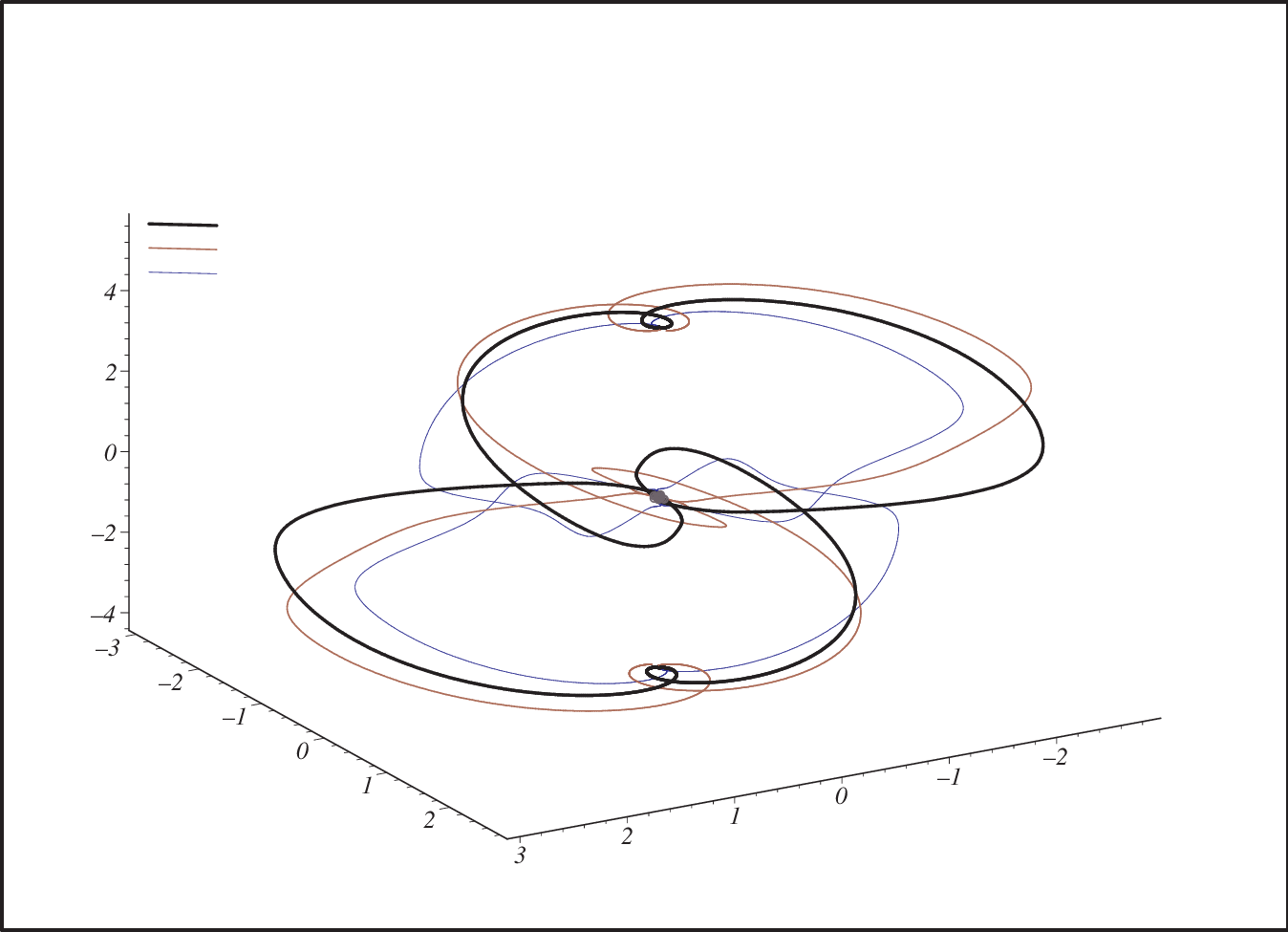}
\put(-192,-50.){\tiny{$z$($\times 10^{-10}$ m)}}
\put(-140,-214){\tiny{$y$($\times 10^{-10}$ m)}}
\put(-37.,-186){\tiny{$x$($\times 10^{-10}$ m)}}
\put(-177,-53) {\tiny $a_r=1,b_r=0,a_{\vartheta}=1,b_{\vartheta}=0,a_{\varphi}=1,b_{\varphi}=0$}
\put(-180,-60) {\tiny $a_r=2.4,b_r=-0.4,a_{\vartheta}=2,b_{\vartheta}=0.5,a_{\varphi}=1.4,b_{\varphi}=0.3$}
\put(-184,-66) {\tiny $a_r=4,b_r=2,a_{\vartheta}=1.25,b_{\vartheta}=0.75,a_{\varphi}=2,b_{\varphi}=-0.8$}
}
\raisebox{-1.25cm}{\centerline{\vbox{ \hsize=7cm\noindent\scriptsize   Fig5.
Quantum 3D spatial trajectories of the Hydrogen's electron
                in the first excited state ($E=-13.6/4$\, eV, $n=2$, $l=0$,
                $m_l=0$).}}}
\end{figure}

\vskip\baselineskip \noindent {\bf b- excited state: {
\mathversion{bold}${n=2}$, ${l=0}$, ${m_l=0}$}
}\vskip0.5\baselineskip
For this state, the radial function ${\cal X}_1^{(20)}$ is
\begin{equation}\label{rad20}
{\cal X}_1^{(20)}(r)=\left({r\over a_o}\right)\left(2-{r\over a_o}\right)\exp\left(-{r\over 2a_o}\right)\, \; ,
\end{equation}
while the angular functions still given by Eqs. (\ref{theta10}) and (\ref{phi10}). Replacing Eq.
(\ref{rad20}) into Eq. (\ref{radmom}), we find
\begin{equation}\label{radmom20}
{dZ\over dr}=\pm {\hbar a_r\left({r\over a_o}\right)^{-2}\left(2-{r\over a_o}\right)^{-2}\, e^{r\over a_o}\over
(a_r I_2(r)+b_r)^2+1}\, \; ,
\end{equation}
where
\begin{eqnarray}
I_2(r)=\, \int \left({a_o\over r}\right)^{2}\left(2-{r\over a_o}\right)^{-2}\, \exp\left({r\over a_o}\right) dr \hskip24mm\nonumber\\
=-{a_o\over 4}{\exp\left({r\over a_o}\right)\over {r\over a_o}-2}-{1\over 4}{a_o^2\over r}\exp\left({r\over a_o}\right)-{a_o\over 2}{\mathrm{Ei}}\left(1,-{r\over a_o}\right)\, , \nonumber
\end{eqnarray}
Taking Eq. (\ref{radmom20}) into Eq. (\ref{quartime}), we find
\begin{equation}\label{radtime20}
{\hbar \over m_0}dt=\,\pm\left(-{r^2\over 4a_o^2}+{2r\over a_o}\right)^{-1}\, \left[{a_r\, a_o^2\left(2-{r\over a_o}\right)^{-2}\, e^{r\over a_o}\over \left(a_r I_2(r)+b_r\right)^2+1}\right]\, dr\; ,
\end{equation}
Integrating numerically Eq. (\ref{radtime20}), we plot the radial
time equation for this bound state (Fig.4). In this figure, the
electron moves between two radial positions $r = 0$ and $r = 8\, ao$
through different trajectories. All these trajectories pass throw
some nodes. The radial positions $r=0$ of the nodes correspond to the
zeros of the radial function ${\mathcal X}_1^{(20)}$. In particular,
the nodes situated at $r = 8\, ao$ correspond to the vanishing value
of the radial velocity given by Eq. (\ref{radtime20}). The radial
velocities do not vanish at $r=0$ making the electron going through
this position toward another extremity $r=8ao$ different from the
previous one in the same way as for the ground state case. Thus,
there are two extremities $r=8ao$ and four trajectories cut branches
($r_{extr1}=8ao\to r=0$; $r=0\to r_{extr2}=8ao$;
$r_{extr2}=8ao\to r=0$ and $r=0\to r_{extr2}=8ao$).

\noindent Taking Eqs. (\ref{radmom20}), (\ref{thetamom10}) and
(\ref{phimom10}) into Eqs. (\ref{quardyn}) and (\ref{quaphidyn}),
we get
\begin{equation}\label{raddyn20}
{4\, a_\varphi\, \sin^2(\vartheta)\,  \over (a_\varphi\, \varphi
+b_\varphi )^2+1}\, d\varphi=\pm{a_r\, a^2_o\left(-{r^2 \over
4a^2_o}+{2 r\over a_o\, }\right)^{-1}\, \left(2-{ r\over a_o\, }\right)^{-2}\, r^{-2}\, e^{r \over a_o}  \over
\left(a_r I_2(r)+b_r\right)^2+1}\, dr\; ,
\end{equation}
and Eq. (\ref{thetadyn10}). Integrating Eqs. (\ref{raddyn20})
and (\ref{thetadyn10}) we plot the 3D quantum spatial trajectories
given in Fig.5 for this state. This figure shows clearly
that for different values of the hidden variables the electron
traces different trajectories that all pass from three nodes situated
at $r=0$ and the two extremities $r=8ao$.

\vskip\baselineskip \noindent {\bf c- excited state: {
\mathversion{bold}${n=2}$, ${l=1}$, ${m_l=1}$}
}\vskip0.5\baselineskip
\begin{figure}\label{time211}
\def\put(#1,#2)#3{\leavevmode\rlap{\hskip#1\unitlength\raise#2\unitlength\hbox{#3}}}
\centerline{\includegraphics[height=7cm,width=7cm,angle=270]{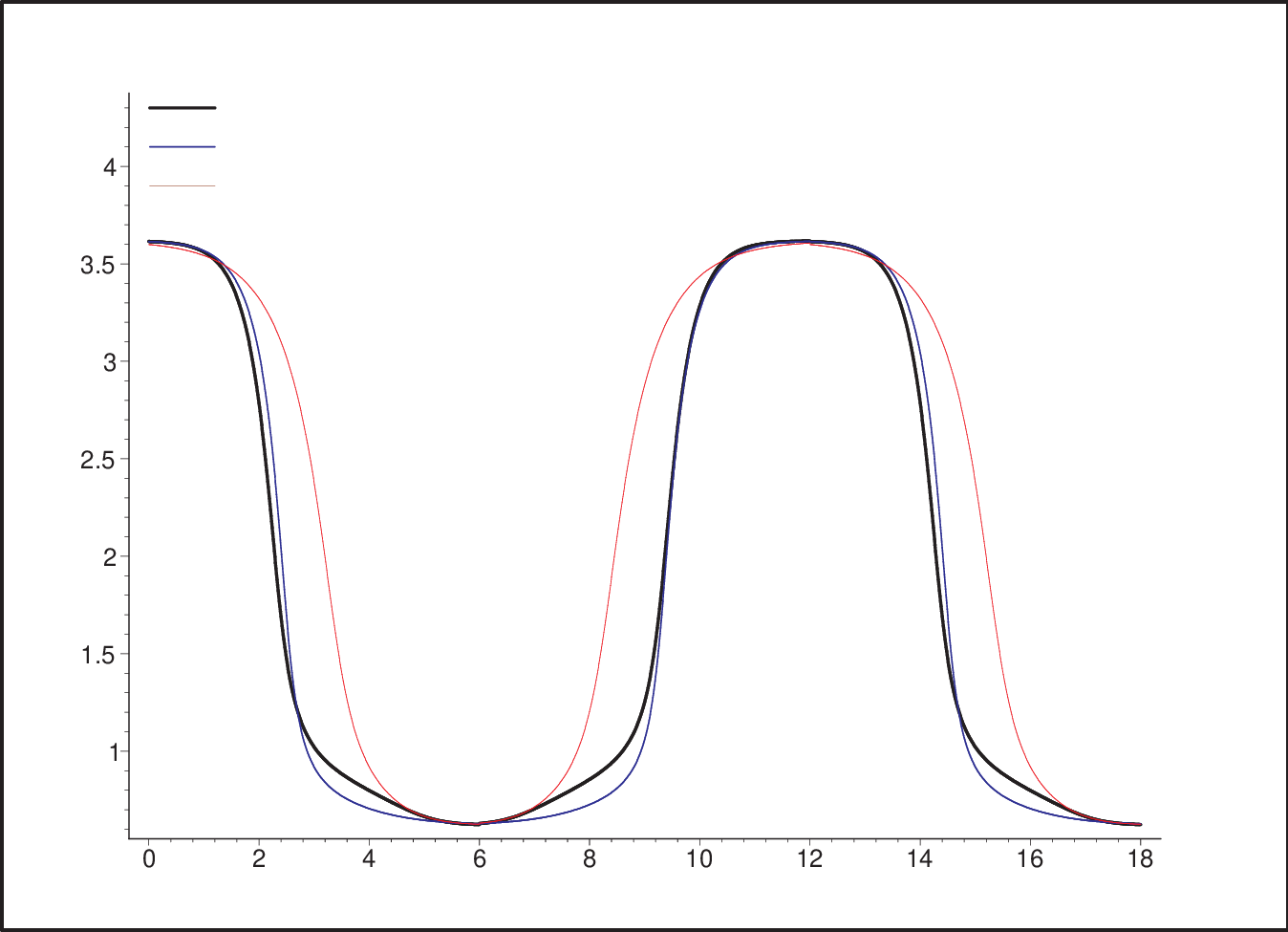}
\put(-35,-188){\tiny{$t$($\times 10^{-16}$ s)}}
\put(-200.,-15){\tiny{$r$($\times 10^{-10}$ m)}}
\put(-173,-23) {\tiny $a_r=7,b_r=2.25$}
\put(-177,-31) {\tiny $a_r=4.5,b_r=2$}
\put(-180,-39) {\tiny $a_r=3,b_r=1$}
}
\raisebox{-1.cm}{\centerline{\vbox{ \hsize=8cm\noindent\scriptsize   Fig6. Quantum radial time trajectories for the Hydrogen's
                electron in the second excited state ($E=-13.6/4$\,
                eV, $n=2$,$l=1$,$m_l=1$). For all the curves, we have
                chosen $r(t=0)= 0$.}}}
\end{figure}
For this state , the radial function is
\begin{equation}\label{X21}
{\cal X}^{(2 1)}_1(r)=\left({r\over a_o}\right)^2\, \exp\left(-{r\over 2a_o}\right)\, \; ,
\end{equation}
and the angular functions are
\begin{equation}\label{theta11}
{\cal T}^{(1\, 1)}_1(\vartheta)=\sin^{3 \over 2}(\vartheta)\, \; ,
\end{equation}
\begin{equation}\label{Phi1}
F^{(1)}_1=\sin(\varphi)\; ,\hskip9mm
\end{equation}
without taking into account of the multiplicative constant $(-2)$
obtained in the calculation of ${\cal T}^{(1\, 1)}_1$. Replacing
Eqs. (\ref{X21}), \ref{theta11} and (\ref{Phi1}) into Eqs. (\ref{radmom}), (\ref{thetamom}) and (\ref{phimom}), we find
\begin{figure}\label{quan211}
\def\put(#1,#2)#3{\leavevmode\rlap{\hskip#1\unitlength\raise#2\unitlength\hbox{#3}}}
\centerline{\includegraphics[height=7cm,width=8cm,angle=270]{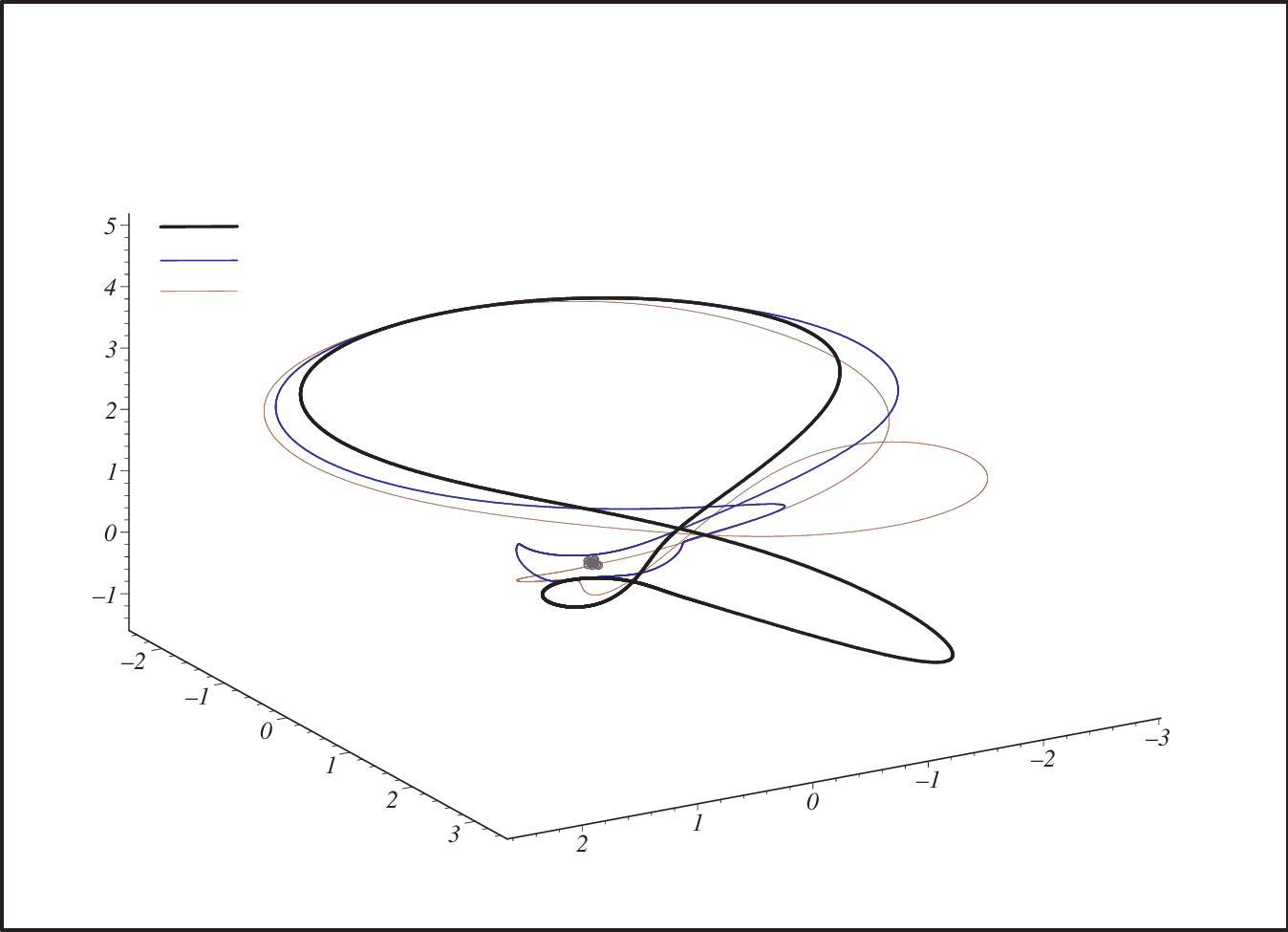}
\put(-192,-50.){\tiny{$z$($\times 10^{-10}$ m)}}
\put(-140,-214){\tiny{$y$($\times 10^{-10}$ m)}}
\put(-37.,-186){\tiny{$x$($\times 10^{-10}$ m)}}
\put(-173,-56) {\tiny $a_r=1,b_r=0,a_{\vartheta}=1,b_{\vartheta}=0,a_{\varphi}=1,b_{\varphi}=0$}
\put(-176,-63) {\tiny $a_r=2,b_r=0,a_{\vartheta}=2,b_{\vartheta}=1,a_{\varphi}=1.5,b_{\varphi}=0.4$}
\put(-180,-70) {\tiny $a_r=10,b_r=-2,a_{\vartheta}=3,b_{\vartheta}=0.75,a_{\varphi}=2,b_{\varphi}=-0.5$}
}
\raisebox{-1.25cm}{\centerline{\vbox{ \hsize=7cm\noindent\scriptsize   Fig7.
Quantum 3D spatial trajectories of the Hydrogen's
                 electron in the third excited state ($E=-13.6/4$\,
                 eV, $n=2$, $l=1$,$m_l=1$).}}}
\end{figure}
%
%
\begin{equation}\label{radmom21}
{dZ \over dr}=
\pm{\hbar a_r\; \left({r \over a_o
}\right)^{-4}e^{r \over a_o } \over (a_r I_3(r)+b_r)^2+1}\; ,
\end{equation}
\begin{equation}\label{thetamom21}
{dL \over d\vartheta}=\pm{\hbar a_\vartheta\;
\sin^{-3}(\vartheta)\,  \over \left[a_\vartheta
\left({\cos(\vartheta) \over \cos(2\, \vartheta)-1}+\, {1 \over
2} \log(|\tan({\theta \over 2})|)\right)
+b_\vartheta\right]^2+1}\; ,
\end{equation}
\begin{equation}\label{phimom21}
{dM \over d\varphi}=\pm\, {\hbar\, a_\varphi \over (a_\varphi\,
\sin(\varphi)+b_\varphi\, \cos(\varphi))^2+\cos(\varphi)^2}  \; ,
\end{equation}
where
\begin{eqnarray}
I_3(r)=\; \int \left({r\over a_o}\right)^{-4}
\exp\left({r\over a_o}\right)\, dr\hskip54mm\nonumber\\
=-{a_o\over 3}\left[{a_0\over r}\exp\left({r\over a_o}\right)
\left[\left({a_o\over r}\right)^2+{1\over 2}{a_o\over r}+{1\over 2}\right]+{1\over 2}\mathrm{Ei}\left(1, -{r\over a_o}\right)\right]\, .\nonumber
\end{eqnarray}
Taking Eq. (\ref{radmom21}) into Eqs. (\ref{quartime}), we find
\begin{equation}\label{radtime21}
{\hbar \over m_0}\; dt=\pm\, {a_r\, a^2_o \left(-{r^2 \over 4\,
a^2_o}+{2 r\over a_o\, }-2\right)^{-1}\, \left({r \over
a_o}\right)^{-2}\, e^{r \over a_o}  \over \left(a_rI_3(r)+b_r\right)^2+1}\; dr\; .\hskip5mm
\end{equation}
Integrating numerically Eq. (\ref{radtime21}), we plot the radial time equation (Fig.6)
for this bound state. The electron moves between two radial extremities
$r=(4-2\sqrt{2})a_o$ and $r=(4+2\sqrt{2})a_o$ constituting nodes for the different
trajectories that the electron traces. At the nodes positions the radial velocity
given by Eq. (\ref{radtime21}) vanishes.

\noindent Taking Eqs. (\ref{radmom21}), (\ref{thetamom21}) and (\ref{phimom21})
into Eqs. (\ref{quardyn}), (\ref{quathetadyn}) and (\ref{quaphidyn}), we get
\begin{eqnarray}\label{quardyn21}
{{4 \over 3}\; a_\varphi\, sin^2(\vartheta)\,  \over (a_\varphi\,
\sin(\varphi) +b_\varphi \cos(\varphi))^2+\cos^2(\varphi)}\,
d\varphi=
\nonumber\hskip45mm \\
\pm\, {a_r\, a^2_o \left(-{r^2 \over 4\, a^2_o}+{2 r\over a_o\,
}-2\right)^{-1}\, \left({r \over a_o}\right)^{-2}\, r^{-2}\, e^{r \over
a_o} \over \left(a_r I_3(r)+b_r\right)^2+1}\, dr\; ,
\end{eqnarray}
\begin{eqnarray}\label{quathetadyn21}
{a_\varphi\, \over (a_\varphi\, \sin(\varphi) +b_\varphi
\cos(\varphi))^2+\cos^2(\varphi)}\, d\varphi =
\nonumber\hskip40mm \\
 \pm{a_\vartheta\; \left({8 \over 3}\,
\sin^2(\vartheta)-1\right)^{-1}\, \sin^{-3}(\vartheta) \over
\left[a_\vartheta \left({\cos(\vartheta) \over \cos(2\,
\vartheta)-1}+{1 \over 2}\, \log(|\tan({\theta \over 2})|)\right)
+b_\vartheta\right]^2+1}\, d\vartheta\; .
\end{eqnarray}
\begin{figure}\label{QUCLAss}
\def\put(#1,#2)#3{\leavevmode\rlap{\hskip#1\unitlength\raise#2\unitlength\hbox{#3}}}
\centerline{\includegraphics[height=7cm,width=8cm,angle=270]{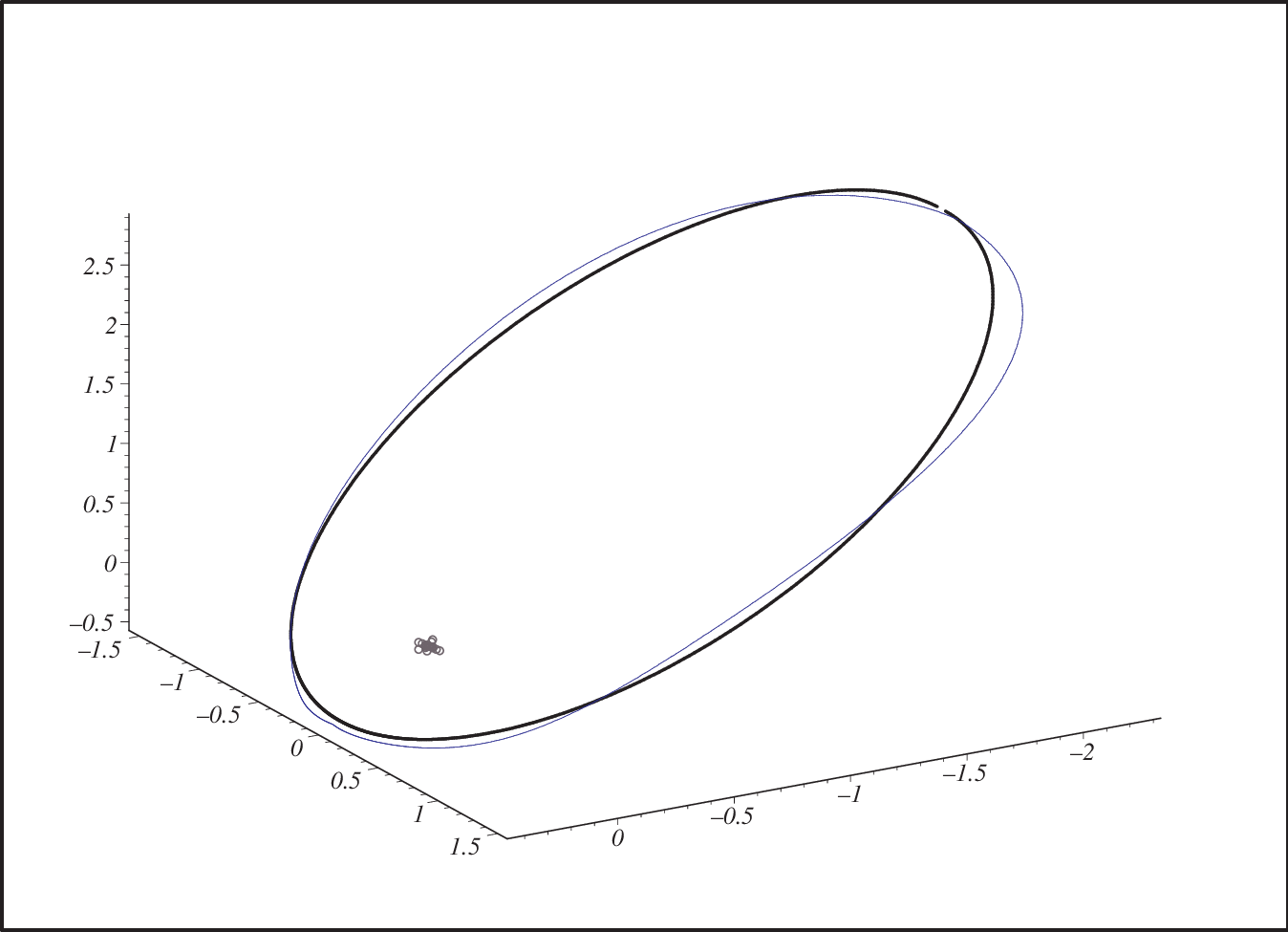}
\put(-192,-50.){\tiny{$z$($\times 10^{-10}$ m)}}
\put(-140,-214){\tiny{$y$($\times 10^{-10}$ m)}}
}
\raisebox{-3.3cm}{\centerline{\vbox{ \hsize=7cm\noindent\scriptsize   Fig8.
 This figure shows a comparison between the purely 3D
                classical trajectory to the one of the possible 3D
                quantum trajectories of the Hydrogen's electron for
                the state ($E=-13.6/4$\, eV, $n=2$,$l=1$, $m_l=1$).
                For the corresponding values of the hidden variables
                ($a_r=3.6$, $b_r=0.1$, $a_\vartheta=1$,$b_\vartheta=0$,
                $a_\varphi=1$, $b_\varphi=0$), we see easily from this picture,
                that the quantum trajectory can be considered as a deformation of
                the elliptic purely classical 3D trajectory.More comparison is
                done in Fig. 9, where projections of the two trajectories into
                $YZ$ and $XZ$ plans are done.}}}
\end{figure}
\noindent Integrating the last equations we plot the 3D quantum spatial
trajectories (Fig.7). For this state, we find that the QTs
are closed curves that are contained between the two
extremities $r=(4-2\, \sqrt{2})\, a_o$ and $r=(4+2\, \sqrt{2})\, a_o$
constituting nodes. Note that, while the classical trajectory is
an ellipse situated into an inclined plan (Figs.1), the QTs
have complex forms and are not contained in a plan (Figs.7 and 8).
Since the radial velocities vanish at the extremities, then,
when the electron reaches the lowest extremity position
$r=r_{min}=(4-2\sqrt{2})a_o$ it must turn back to the highest
extremity position $r=r_{max}=(4+2\sqrt{2})a_o$ making the number
of the trajectory cut branches two. The reason of this behavior is that the electron
should turn around the nucleus making it trajectories to surround it.

\noindent Now, let us compare one specific QT integrated
from Eqs. (\ref{quardyn21}) and (\ref{quathetadyn21}) with
the corresponding classical trajectory plotted from the
classical equations (\ref{clasrdyn}) and (\ref{clasphidyn}).
Such comparison can be done only for the states for which
$m_l\neq 0$, otherwise the classical conjugate momentum $dM/d\varphi$
with respect to $\varphi$ takes an imaginary value.
This is the reason why we considered the state for which
$n=2$, $l=1$ and $m_l=1$. For this state, the corresponding constants
$\alpha$ and $\beta$ take the values $2\hbar^2$ and $(\sqrt{3}/2)\hbar$
respectively. Fig.8 shows that for $a_r=3.6$, $b_r=0.1$,
$a_\vartheta=1$, $b_\vartheta=0$, $a_\varphi=1$ and $b_\varphi=0$,
the QT goes approximately to the classical trajectory with an apparent
residual deformation (Figs.8 and 9) that we assign
to the residual quantum effects. To render this note more
clear, we plotted in Fig.9 the projection of the two
trajectories into $XZ$ and $YZ$ plans. The Fig.9 shows
that the QT has a deformation of an ellipse.

\medskip

It is useful to notice that for the different bound state studied above,
the electron is trapped between two radial positions
given by the two roots of the equation
\begin{equation}\label{rootseq}
E-V(r)-{l(l+1)\, \hbar^2 \over 2\, m_0\, r^2}=0\, ,
\end{equation}
\begin{figure}\label{comparison}
\def\put(#1,#2)#3{\leavevmode\rlap{\hskip#1\unitlength\raise#2\unitlength\hbox{#3}}}
\centerline{\includegraphics[height=4.6cm,width=6.5cm,angle=270]{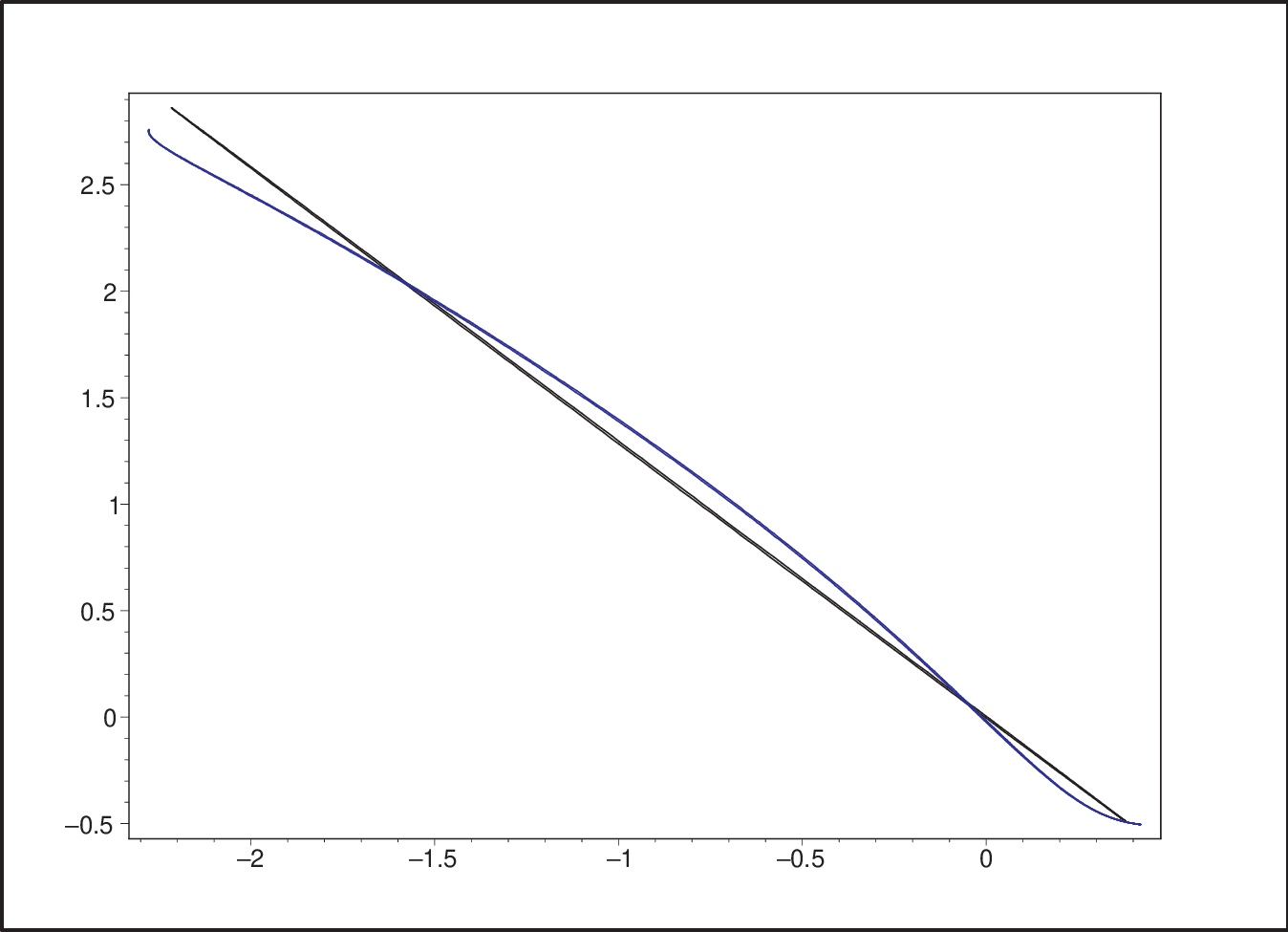}
\put(-30,-174){\tiny{$x$($\times 10^{-10}$ m)}}
\put(-135.,-15){\tiny{$z$($\times 10^{-10}$ m)}}
\includegraphics[height=4.6cm,width=6.5cm,angle=270]{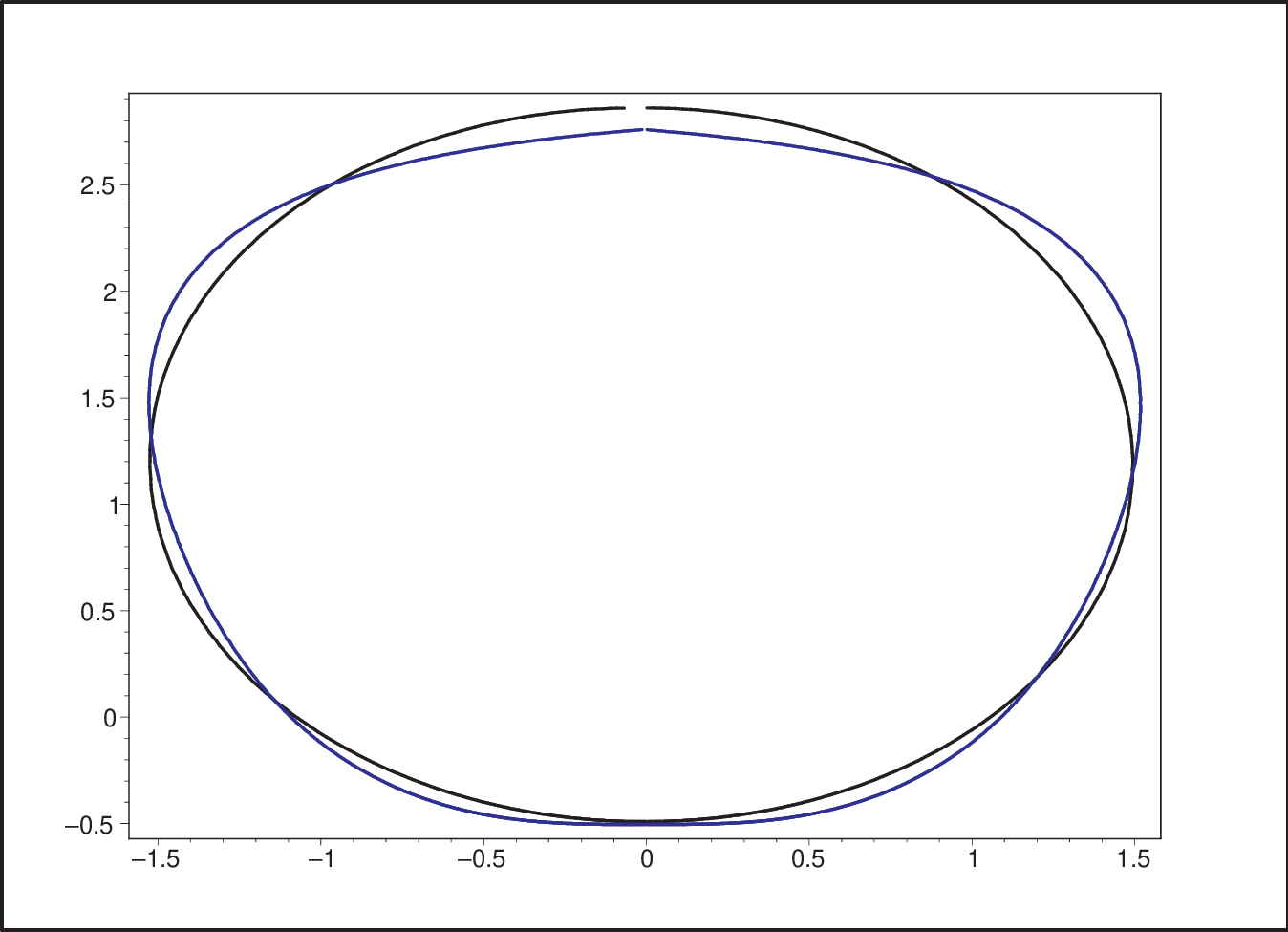}
\put(-30,-174){\tiny{$y$($\times 10^{-10}$ m)}}
\put(-135.,-15){\tiny{$z$($\times 10^{-10}$ m)}} }
\raisebox{-1.5cm}{\centerline{\vbox{ \hsize=8cm\noindent\scriptsize   Fig9. Projection onto the $XZ$ and $YZ$ plans of the spatial
                trajectory for the state ($E=-13.6/4$\, eV, $n=2$, $l=1$,
                $m_l=1$)($a_r=3.6$, $b_r=0.1$,$a_\vartheta=1$,
                $b_\vartheta=0$, $a_\varphi=1$,$b_\varphi=0$), and its
                corresponding purely classical trajectory. The quantum
                trajectory seems to be a deformation of the classical one.}}}
\end{figure}


\noindent and, if the electron reaches an extremity position in the trapping zone
delimited by the two roots, either it will return back on it way toward
the other extremity or it will be pushed away from the trapping zone.
We think that when the electron is pushed away from the trapping zone,
it will either take a trajectory of another state (excited or fundamental)
or it will be totally ejected from the atom. This fact is partially
established in Fig.10 which shows that for the bound state
($n=1$, $l=0$, $m_l=0$), after crossing the position $r=2\, a_0$, which
is the upper extremity of the trapping zone,
the electron reaches in a limited time a position of about $10^{10}$ m.
This means that, after living the trapping zone the electron will be
strongly ejected far from the atom and is no more bounded to it.
That is in some disagreement with the standard quantum mechanics which
suggests that for any position in the space, the atomic electron has
a non vanishing probability to be bounded to the atom. A consequence of
such a result is a partial localization of the electron in the space
in the case of the bound states.

\vskip\baselineskip \noindent {\bf 6- Conclusion}
\vskip0.5\baselineskip
First, we emphasize that in this article we
continued the work presented in Ref. \cite{Dja1} where
we have established the 3D quantum law of motion that we apply in this paper
to the central potential case and in particular to the Hydrogen's atom. We
proceeded by three main steps, the first one is the establishment of
the 3D-QSHJE and the quantum reduced action for the case of a
central potential (Sec. 2). Then, in the second step, we
established the 3D Quantum law of motion under the central
potential (Sec. 3). Finally, in the third step, we plotted the 3D
QTs of the Hydrogen's electron (Sec. 5).
Analyzing the trajectories obtained in
Sec. 5 for different bound states of the electron, we found
three main outcomes.
\vskip0.3\baselineskip
\begin{figure}\label{eleject}
\def\put(#1,#2)#3{\leavevmode\rlap{\hskip#1\unitlength\raise#2\unitlength\hbox{#3}}}
\centerline{\includegraphics[height=6cm,width=7cm,angle=270]{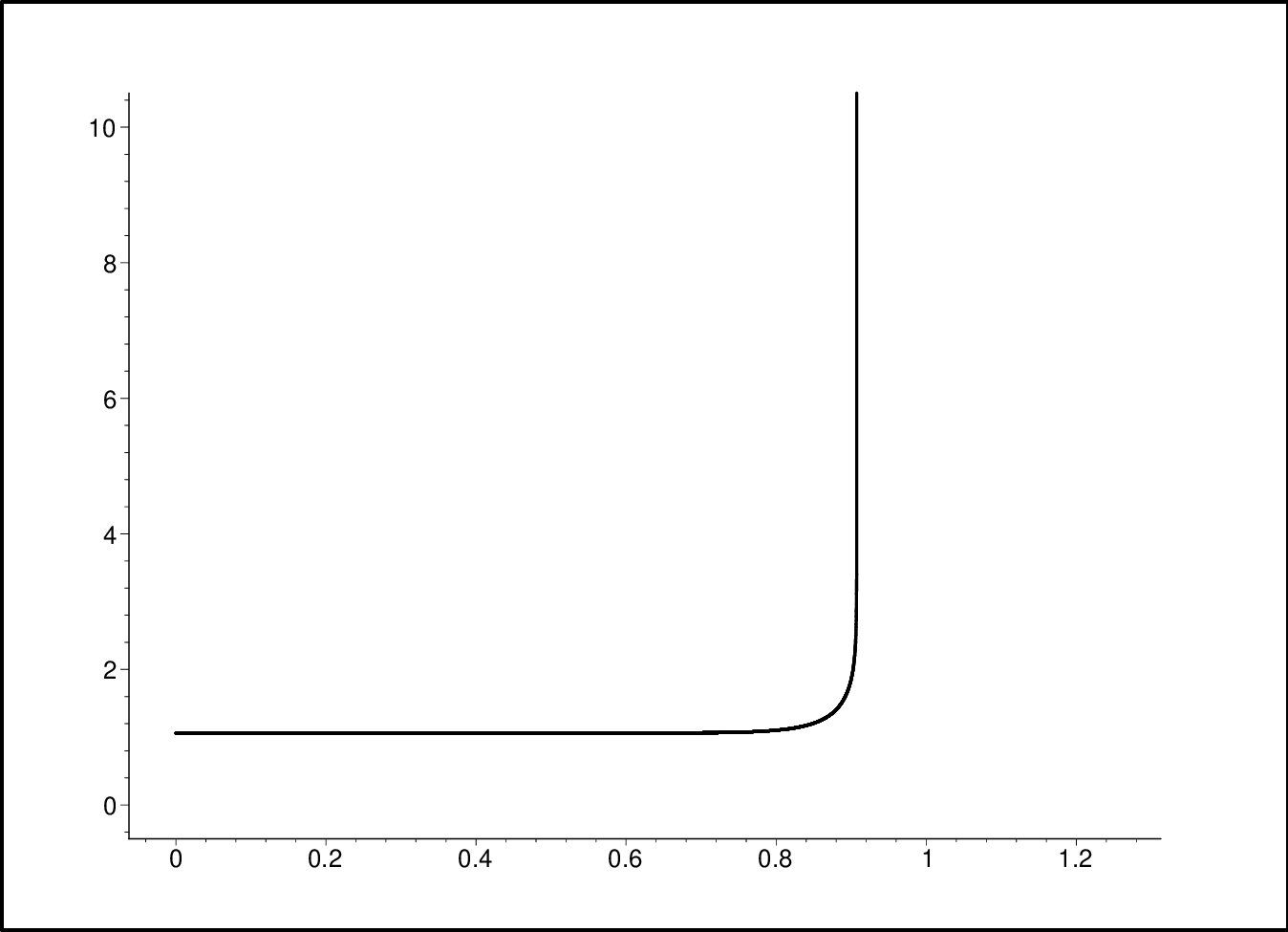}
\put(-26,-185){\tiny{$t$($\times 10^{-16}$ s)}}
\put(-185.,-18){\tiny{$r$($\times 10^{-10}$ m)}} }
\raisebox{-1.35cm}{\centerline{\vbox{ \hsize=8cm\noindent\scriptsize   Fig10. The ejection of
the electron from the atom when it leave the trapping zone $(0\leq
r \leq 2\, a_o)$ in the case of the fundamental bound state
($E=-13.6$\, eV, $n=1$, $l=0$, $m_l=0$). We chose for the radial
hidden variables the values $a_r=1.75$ and $b_r=0.5$.}}}
\end{figure}
\noindent The first outcome is the fact that for the bound state cases,
the electron is trapped between two radial positions given by the two
roots of the Eq.(\ref{rootseq}). A consequence of such a result is a
partial localization of the electron in the space in the case of the
bound states contradicting the standard quantum mechanics outcome which
stipulates that for a giving bound state, the electron is localized
in all the space.

\noindent The second outcome is the existence of nodes from which all the
possible QTs plotted for the same state pass. In general,
the positions of these nodes correspond to the zeros
of the functions ${\cal X}^{(nl)}$ and to the points where
the radial velocity ${\dot r}$ vanishes.

\noindent The third outcome concerns the fact that the QTs goes approximately to its
corresponding purely classical trajectory for well chosen values the hidden variables. That
reinforce the fact that the quantum equations of motion reduce to the classical ones.

\vskip2\baselineskip \noindent {\bf REFERENCES}


\end {document}